\pdfoutput=1

\documentclass[12pt]{article}
\usepackage{amsfonts}
\usepackage{amsmath}
\usepackage{amssymb}
\usepackage{bigints}
\usepackage{booktabs}
\usepackage[nosort]{cite}
\usepackage{color}
\usepackage{dsfont}
\usepackage{float}
\usepackage{framed}
\usepackage{graphicx}
\usepackage{indentfirst}
\usepackage{mathrsfs}
\usepackage{multirow}
\usepackage{pdflscape}
\usepackage{setspace}
\usepackage{subdepth}
\usepackage{subfig}
\usepackage{titlesec}
\usepackage[dotinlabels]{titletoc}
\usepackage{wrapfig}
\usepackage[all]{xy}
\usepackage{young}
\usepackage[vcentermath]{youngtab}
\usepackage{relsize}
\usepackage{stackengine}
\usepackage{datetime}

\usepackage{hyperref}

\numberwithin{equation}{section}

\usepackage{verbatim}

\newcommand{\be}{\begin{equation}}
\newcommand{\ee}{\end{equation}}

\def\({\left(} \def\){\right)}
\def\[{\left[} \def\]{\right]}

\def\sgn{\text{sgn}}

\def\mF{\mathcal{F}}

\def\mO{\mathcal{O}}
\def\mC{\mathcal{C}}

\def\mI{\mathcal{I}}

\def\eps{\epsilon}

\newcommand{\bea}{\begin{eqnarray}}
\newcommand{\eea}{\end{eqnarray}}

\newcommand{\bml}{\begin{multline}}
\newcommand{\emll}{\end{multline}}

\usepackage[left=2.5cm,right=2.5cm,top=2.5cm,bottom=2.5cm]{geometry}
 \usepackage[font={footnotesize}]{caption}

\titleformat{\section}{\normalfont\bfseries}{\thesection.}{4pt}{}
\titlespacing{\section}{0pt}{22pt}{6pt}

\titleformat{\subsection}{\normalfont\itshape}{\thesubsection.}{4pt}{}
\titlespacing{\subsection}{0pt}{18pt}{6pt}

\titleformat{\subsubsection}{\normalfont\itshape}{\thesubsubsection.}{4pt}{}
\titlespacing{\subsubsection}{0pt}{16pt}{6pt}

\def\1{{\mathds 1}}

\DeclareFontShape{OT1}{cmr}{mx}{n}%
    {<->cmr10}{}
\newcommand{\mytitlefont}{\fontseries{mx}\selectfont}
\DeclareMathAlphabet{\titlemath}{OT1}{cmr}{mx}{n}

\usepackage{footmisc}

\begin{document}


\begin{titlepage}

\begin{center}

~\\[2cm]

{\fontsize{20pt}{0pt} \mytitlefont An introduction to the  SYK model }

~\\[0.5cm]

{\fontsize{14pt}{0pt} Vladimir Rosenhaus
}

~\\[0.1cm]

\it{Kavli Institute for Theoretical Physics}\\ \it{University of California, Santa Barbara, CA 93106}

~\\[0.8cm]

\end{center}

\noindent 
These notes are a short introduction to the Sachdev-Ye-Kitaev model. We discuss: SYK and tensor models as a new class of large $N$ quantum field theories, the near-conformal invariance in the infrared, the computation of correlation functions, generalizations of SYK, and applications to  AdS/CFT and  strange metals. 
\vspace{2cm}
\center \textit{In memory of Joe Polchinski}
\vfill



\end{titlepage}

\tableofcontents

\section{Introduction}
The Sachdev-Ye-Kitaev model \cite{SY, Kitaev} is a strongly coupled, quantum many-body system that is chaotic, nearly conformally invariant, and exactly solvable. This remarkable and, to date, unique combination of properties have driven the intense activity surrounding SYK and its applications within both high energy and condensed matter physics. 

As a quantum field theory, SYK and, more generally, tensor models, constitute a new class of large $N$ theories. The dominance of a simple and well-organized set of Feynman diagrams, iterations of melons, enables the computation of all correlation functions. As a solvable model of holographic duality, SYK accurately captures two-dimensional gravity, and has  the potential to shed light on the workings of holography and black holes. As a solvable many-body system, SYK serves as a building block  for constructing  a metal, capturing some of the properties of non-Fermi liquids. 

These notes are a brief introduction to SYK. 
In Sec.~\ref{sec:N} we review large $N$ field theories, in particular vector models, and introduce SYK and tensor models. In Sec.~\ref{Sec:Infrared} we discuss the low energy limit of SYK,  described by a sum of the Schwarzian action and a conformally invariant action. In Sec.~\ref{sec:correlation} we discuss how  the simple Feynman diagrammatics of large $N$ SYK, combined with the power of conformal symmetry,  allows for an explicit computation of all correlation functions. In Sec.~\ref{sec:applications} we discuss applications of SYK to holographic duality, and to strange metals. 

\section{A New Large $N$ Limit} \label{sec:N}

Large $N$ quantum field theories are theories with a large number of fields, related by some symmetry, such as $O(N)$. Their essential property is the factorization of  correlation functions of $O(N)$ invariant operators, $\langle \mO(x_1) \mO(x_2)\rangle = \langle \mO(x_1) \rangle\langle \mO(x_2)\rangle + \frac{1}{N}\( \ldots \)$. As such, large $N$ theories are in a sense semi-classical, with $1/N$ playing the role of $\hbar$.

\subsection{Vector models}
\begin{figure}
\centering
\includegraphics[width=.8in]{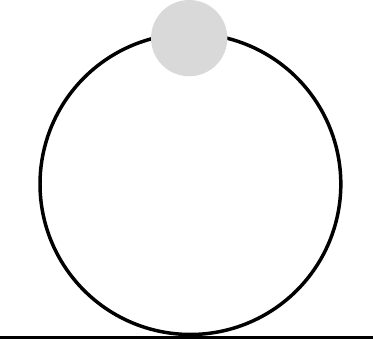}
\caption{The self-energy for the $O(N)$ vector model, in terms of the propagator. Iterating gives a sum of bubble diagrams.} \label{FigBubble}
\end{figure}

The simplest large $N$ quantum field theories are vector models \cite{Coleman, Gross:1974jv}. An example is the   $O(N)$ vector model, having $N$ scalar fields, $\vec\phi = (\phi^1, \ldots, \phi^N)$, with a quartic $O(N)$ invariant  interaction, 
\be
I = \int d^d x \( \frac{1}{2} (\partial \vec{\phi})^2 + \frac{1}{2} \mu^2 \vec{\phi}^{\,2} + \frac{g}{4} ( \vec{\phi} \cdot \vec{\phi})^2 \)~.
\ee
In dimensions $2<d<4$, if one appropriately tunes the bare mass $\mu$,  there is an infrared fixed point: the Wilson-Fisher fixed point, describing magnets.

The power of large $N$ is that instead of studying the theory perturbatively in the coupling $g$,  one can reorganize the perturbative expansion, into powers of $g N$ and $1/N$. At a given order in $1/N$, one is able to compute to all order in $g N$. 
For instance, at leading order in $1/N$, the only Feynman diagrams contributing to the two-point function of the $O(N)$ vector model are bubbles, see Fig.~\ref{FigBubble}, all of which are summed by the  integral equation, 
\be \label{eq:1}
G(p) = \frac{1}{p^2 + \mu^2 + \Sigma(p)}~, \ \ \ \ \Sigma(p)  =  g N \int d^d q\, G(q)~,
\ee
where $G(p)$ is the momentum space two-point function, and $\Sigma(p)$ is the self-energy. 
The self-energy is independent of the momentum: the only effect of the bubble diagrams is to shift the mass. Defining $m^2 = \mu^2 + \Sigma$, the above Schwinger-Dyson equation becomes, 
\be \label{eq:m}
m^2 = \mu^2 + g N \int d^d q \frac{1}{q^2 + m^2}~.
\ee

An equivalent way of analyzing the $O(N)$ vector model is by introducing an auxiliary Hubbard-Stratonovich field $\sigma$, so as to rewrite the action as, 
\be \label{eq:HS}
I = \int d^d x \(  \frac{1}{2} (\partial \vec \phi)^2 + \frac{1}{2}\mu^2 \vec \phi^{\, 2} + \frac{1}{2} \sigma \vec \phi^{\, 2} - \frac{\sigma^2}{4 g}\)~.
\ee
Integrating out $\sigma$ gives back the original action. Alternatively, integrating  out the $\phi^a$ fields, gives an action involving only $\sigma$,
\be
\frac{I_{\sigma}}{N} =\frac{1}{2} \log \det(- \partial^2 + \mu^2 + \sigma) - \frac{1}{4g N}\int d^d x \, \sigma^2~.
\ee
The saddle of $I_{\sigma}$ gives back the Schwinger-Dyson equation for summing bubbles. More generally, one could have introduced a source for $\phi$,
and then used the resulting $I_{\sigma}$ to, in principle, compute any correlation function of $O(N)$ invariant operators, to any order in $1/N$.

\subsubsection*{Matrix Models}
There are many systems, such as  Yang-Mills theory, in which the fundamental fields are matrices, rather than vectors.~\footnote{There are also large $N$ models involving both vectors and matrices, such as, in one dimension, the Iuzika-Polchinksi model \cite{Iizuka:2008hg}, in two dimensions,  the 't Hooft model of QCD \cite{tHooft:1974pnl, Callan:1975ps}, and in three dimensions, Chern-Simons theory  coupled to matter \cite{Giombi:2011kc, Aharony:2011jz}. These models are vector-like, as the matrix degrees of freedom have no self-interactions, and are solvable at large $N$ through summation of rainbow diagrams. } The large $N$ dominant Feynman diagrams in such theories are planar, when drawn in double line notation \cite{tHooft:1973alw}. There is no known way of summing all planar diagrams, and matrix models are in general not solvable. For some special theories, there are alternate techniques. For instance, models of a single matrix in zero and one dimension can be solved through a map to free fermions \cite{DiFrancesco:1993cyw, Klebanov:1991qa }. More recently, powerful integrability techniques have been applied to planar $\mathcal{N}=4$  super Yang-Mills in four dimensions\cite{Beisert:2010jr}, yielding, for instance,  exact results for  anomalous dimensions \cite{Gromov:2013pga} and progress for the three-point function \cite{Basso:2015zoa}. See \cite{Zamolodchikov:1978xm} for an introduction to integrability, in two dimensions, and \cite{Minahan:2002ve} for the initial discovery of the link between the computation of anomalous dimensions and the diagonalization, via Bethe ansatz, of certain integrable spin chain Hamiltonians.

\subsection{Tensor models and SYK}
\begin{figure}
\centering
\includegraphics[width=1in]{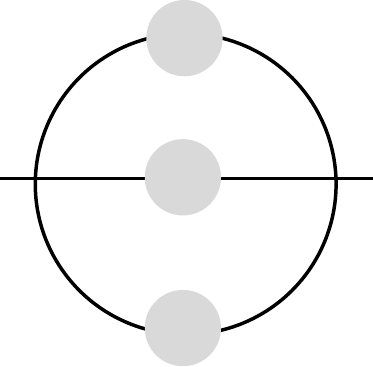}
\caption{The self-energy for SYK and tensor models, in terms of the propagator. Iterating gives a sum of melon diagrams.} \label{FigMelon}
\end{figure}

Having discussed vector models and matrix models, it is  natural to consider tensor models, with fields having three or more indices.
An example of such a model, for  fermions in one dimension and rank-3 tensors, is the Klebanov-Tarnopolsky model \cite{Klebanov:2016xxf}, a simplification of   the Gurau-Witten model \cite{Gurau:2009tw,Witten:2016iux} (see also \cite{Bonzom:2011zz,2016LMaPh.106.1531C,  Delporte:2018iyf, Klebanov:2017nlk}), with the Lagrangian,
\be \label{KT}
L = \frac{1}{2} \sum_{a, b, c =1}^N \psi_{a b c} \partial_{\tau} \psi_{a b c } - \frac{g}{4} \sum_{a_1, \ldots, c_2 = 1}^N  \psi_{a_1 b_1 c_1} \psi_{a_1 b_2 c_2} \psi_{a_2 b_1 c_2} \psi_{a_2 b_2 c_1}~,
\ee
where the real field $\psi_{a b c}$ transforms in the tri-fundamental representation of  $O(N)^3$. Remarkably, the two-point function is dominated by melon diagrams in the limit of large $N$ and fixed $J^2\equiv g^2 N^3$. The summation of all melon diagrams is encoded in the Schwinger-Dyson equation,  see Fig.~\ref{FigMelon},  
\be \label{SD}
G(\omega)^{-1} = - i\omega - \Sigma(\omega)~, \ \ \ \ \ \Sigma(\tau) = J^2 G(\tau)^{3}~,
\ee
where $G(\omega)$ is the Fourier transform of $G(\tau)$. 
Tensor models, which sum melon diagrams, are more rich, and more difficult, than vector models. They are however, at least in some ways, simpler than matrix models, for which there is no closed set of equations to sum the planar diagrams. 

One challenge in the study of tensor models, beyond the melonic dominance at large $N$, is the relatively low degree of symmetry: the number of degrees of freedom scales as $N^3$, while the rank of the symmetry group, $O(N)^3$, scales as $N^2$. 
This difficulty is alleviated by the SYK model \cite{Kitaev}, at the expense of introducing disorder,
\be
L=\frac{1}{2}\sum_{i=1}^N \chi_i \partial_{\tau} \chi_i -\frac{1}{4!} \sum_{i, j, k, l=1}^N J_{i j k l} \chi_i \chi_j \chi_k \chi_l~, \label{SYKL}
\ee
where the couplings are Gaussian-random, with zero mean, and variance $\langle J_{i j k l } J_{i j k l}\rangle  = 6 J^2/N^3$. 
The leading large $N$ diagrams in SYK are melons; identical to those in the tensor model.~\footnote{In fact, the melonic dominance in SYK can be proved by viewing it as a kind of tensor model, with $J_{i j kl}$ being the tensor field \cite{Gurau:2017xhf}, although this proof is  more involved than the standard one for SYK, involving the effective action (\ref{Ieff}). } The subleading $1/N$ corrections, as well as the symmetry group, are however different. 
Taking the partition function, and disorder averaging/integrating out the $J_{i j k l}$
gives a bilocal, $O(N)$ invariant action.~\footnote{The SYK model has quenched disorder. For many quantities, the model is self-averaging, and computing with some randomly chosen, but fixed, couplings should give the same result as averaging over the couplings. In this sense, the disorder average is a trick, in order to be able to analytically perform the calculation. Alternatively, one might wish to view the couplings $J_{i j k l}$ as quantum scalar fields with a two-point function that is a constant (annealed disorder). In fact, up to  order  $1/N^3$, this gives the same results as with quenched disorder \cite{MPRS}, assuming there is no replica symmetry breaking, which is implicit in (\ref{Ieff}) where we  dropped the replica indices (the replica off-diagonal terms are subleading \cite{Kitaev:2017awl}).}
 Introducing bilocal fields $\Sigma(\tau_1, \tau_2)$ and $G(\tau_1, \tau_2)$, with $\Sigma$ acting as a Lagrange multiplier field enforcing  $G(\tau_1, \tau_2) = \frac{1}{N}\sum_{i=1}^N\chi_i(\tau_1) \chi_i(\tau_2)$, and integrating out the fermions, one is left with an action for $\Sigma$ and $G$ \cite{Kitaev}, see also \cite{MS, GR, Kitaev:2017awl},
\be \label{Ieff}
\frac{I_{eff}}{N} = - \frac{1}{2} \log \det\( \partial_{\tau}  - \Sigma\) + \frac{1}{2} \int d\tau_1 d\tau_2 \( \Sigma(\tau_1, \tau_2)G(\tau_1, \tau_2) - \frac{J^2}{4}G(\tau_1, \tau_2)^4\)~.
\ee

Compared with the $O(N)$ vector model, which was captured by an action for a local field $\sigma(x)$, SYK is instead captured by the above action for bilocal fields $\Sigma(\tau_1, \tau_2)$ and $G(\tau_1, \tau_2)$.~\footnote{The solution of the $O(N)$ vector model is an example of mean field theory, whereas  the solution of SYK is an example of dynamical mean field theory \cite{DMFT}.} With this action, the model is in principle solved: instead of the original $N$ fields, there are now only two fields. At infinite $N$, the theory is classical, with the path integral  dominated by the saddle  for $\Sigma$ and $G$, given by (\ref{SD}), reflecting the summation of melon diagrams. All higher point correlation functions follow from expanding the action in powers of $1/N$. The rest of the notes are devoted to computing these in an explicit form, and understanding their physical consequences.

\section{The Infrared} \label{Sec:Infrared}
 In this section we study SYK in the infrared limit,
 following \cite{Kitaev:2017awl}. 
We take the effective action (\ref{Ieff}), and for convenience define, $\sigma(\tau_1, \tau_2) =  \delta(\tau_1 - \tau_2)\partial_{\tau}$, and change variables  $\Sigma \rightarrow \Sigma + \sigma$, so that the  action becomes, $I_{eff} = I_{CFT} + I_{S}$, where,
\bea
\frac{I_{CFT}}{N}&=&- \frac{1}{2} \log \det( - \Sigma) + \frac{1}{2} \int d\tau_1 d\tau_2 \( \Sigma(\tau_1, \tau_2)G(\tau_1, \tau_2) - \frac{J^2}{4}G(\tau_1, \tau_2)^4\)\\
 \frac{I_{S}}{N}& =&\frac{1}{2}\int d\tau_1 d\tau_2\,  \sigma(\tau_1, \tau_2)G(\tau_1, \tau_2)~.
\eea
In the infrared, $|J\tau|\gg 1$, at leading order, we simply drop the $I_S$ part of the action, as the delta function in $\sigma$ is a very UV term. The saddle of $I_{CFT}$ is the Schwinger-Dyson equation from  before, without the $\partial_{\tau}$ term, and its  solution takes the form of a conformal field theory two-point function \cite{SY}, 
\be \label{2pt}
G(\tau_1, \tau_2) =b\frac{\sgn(\tau_{12})}{|J \tau_{12}|^{2\Delta}}~, \ \ \ \text{where}\ \ \  b^4 = \frac{1}{4\pi},\ \ \ \  \Delta = \frac{1}{4},\ \ \ \ \tau_{ij} \equiv \tau_i - \tau_j~.
\ee
One might assume that any higher point correlation function, computed using $I_{CFT}$, would also be conformally invariant. In fact, this is almost true, but not completely. 
Notice that $I_{CFT}$ is time reparametrization invariant, $\tau\rightarrow f(\tau)$, provided $G$ and $\Sigma$ transform appropriately, 
\be \nonumber
G(\tau_1, \tau_2) \rightarrow f'(\tau_1)^{\Delta} f'(\tau_2)^{\Delta}G(f(\tau_1), f(\tau_2))~, \ \ \  \ \ \Sigma(\tau_1, \tau_2) \rightarrow f'(\tau_1)^{1-\Delta} f'(\tau_2)^{1-\Delta} \Sigma(f(\tau_1), f(\tau_2))~.
\ee
As a result, in addition to the solution (\ref{2pt}), we have an entire space of solutions \cite{Kitaev},
\be \label{2pt2}
G(\tau_1, \tau_2) = b\,\frac{ \sgn(\tau_{12})}{J^{2\Delta}} \frac{f'(\tau_1)^{\Delta} f'(\tau_2)^{\Delta}}{|f(\tau_1) - f(\tau_2)|^{2\Delta}}~,
\ee
 and moving between them has no action cost.
At a practical level, this means that using $I_{CFT}$ to compute correlation function will lead to divergences \cite{Kitaev, PR, MS}. Of course, in the full action, with $I_S$ included, there is a cost for $\tau\rightarrow f(\tau)$, so this simply means that we need to move slightly away from the deep infrared limit: rather than just dropping $I_S$, we should view it as a perturbation of the infrared action, and compute its effect, to leading order. We approximate $I_S$ by inserting for $G$ the saddle (\ref{2pt2}). Since $\sigma(\tau_1, \tau_2)$ is a delta function, the integral in the action picks out the $\tau_{12}\ll 1$ part of $G$. Taylor expanding $G(\tau_1, \tau_2) $ about $\tau_+ = (\tau_1 + \tau_2)/2$, for $\tau_{12}\ll 1$,
\be\nonumber
G \rightarrow b\frac{\sgn(\tau_{12})}{|J \tau_{12}|^{2\Delta}} \(1 + \frac{\Delta}{6} \tau_{12}^2\, \text{Sch}(f(\tau_+), \tau_+)+ \ldots\)~, \ \ \ \ \text{where } \ \ \ \text{Sch} (f(\tau), \tau) = \frac{f'''}{f'} - \frac{3}{2}\(\frac{f''}{f'}\)^2
\ee
and we get that, 
\be
\frac{I_S}{N} = \frac{\#}{J} \int d\tau\,  \text{Sch} (f(\tau), \tau) + \ldots
\ee
where $\text{Sch} (f(\tau), \tau)$ is the Schwarzian. The prefactor can not be fixed by this procedure; it must be determined numerically \cite{MS,Kitaev:2017awl}, from the exact solution to the Schwinger-Dyson equation (\ref{SD}). The reason is that we are studying the infrared action, valid for $|J \tau|\gg 1$, while the perturbation $\sigma$ involves a delta function of time, outside the domain of validity of perturbation theory. 

The field $f(\tau)$ is sometimes referred to as the reparametrization mode, or the soft mode, or the $h=2$ mode, or the gravitational mode.~\footnote{See footnote \footref{4ptFoot}, and the start of Sec.~\ref{adscft}, respectively, in order to understand the latter two names.} It is the Nambu-Goldstone mode for the breaking of time reparametrization invariance \cite{MSY}. For further studies of the Schwarzian, see \cite{ Jevicki:2016bwu,  Bagrets:2016cdf, Stanford:2017thb, Mertens:2017mtv,Mertens:2018fds,Mandal:2017thl, Alekseev:2018pbv, Belokurov:2018aol }, as well as references in Sec.~\ref{adscft}, regarding dilaton gravity. 

\vspace{-.19cm}
\section{Correlation Functions} \label{sec:correlation}
We now turn to higher point correlation functions, following the discussion in \cite{GR4}. The four-point function, like the two-point function, is given by the solution of an integral equation, while even higher point functions are given by integrals of products of four-point functions. 
 In the infrared, where there is  near-conformal symmetry, we can go further and write explicit expressions for the higher point correlation functions.~\footnote{
The near-conformal symmetry means that the functional form of correlation functions will have, in addition to the conformal contributions, some pieces that involve mixing with the $h=2$ mode.   These are clearly distinguished from the conformal pieces, as they come with extra factors of $J$. Alternatively, there is a variant of SYK, cSYK \cite{GR3}, which is fully conformally invariant. See also footnote \footref{4ptFoot}. We will only discuss the conformal contributions to correlation functions; it is straightforward to include the others.} 
\vspace{-.17cm}
\subsection{Conformal blocks}
We first recall the constraints that conformal symmetry places on the form of correlation functions. 
A one-dimensional conformal field theory, CFT$_1$,  has  $SL_2(R)$ symmetry,
\vspace{-.12cm}
\be
\tau \rightarrow \frac{a\tau + b}{c\tau + d}~, \ \ \ \ ad-b c=1~.
\ee
Any such transformation is generated by a combination of 
translations $\tau\rightarrow \tau + a$, dilatations $\tau \rightarrow \lambda \tau$, and inversions $\tau\rightarrow \frac{1}{\tau}$. The symmetry fully fixes the functional form of the two-point and three-point functions, 
\vspace{-.1cm}
\be
\langle \mO_h(\tau_1) \mO_h(\tau_2) \rangle = \frac{1}{|\tau_{12}|^{2h}}~, \ \ \ \ \ \ \ \ \ 
\langle \mO_1 \mO_2 \mO_3\rangle = \frac{C_{h_1 h_2 h_3}}{|\tau_{12}|^{h_1+h_2-h_3} |\tau_{13}|^{h_1+h_3 - h_2} |\tau_{23}|^{h_2 + h_3 - h_1}}~, \label{3pt}
\ee
where the $\mO_i$, shorthand for $\mO_{h_i}(\tau_i)$, are primary operators of dimension $h_i$, and the $C_{h_1 h_2 h_3}$ are  structure constants. 

To find the functional form of a four-point function, we combine two three-point functions, and integrate over one of the points, 
\vspace{-.2cm}
\be \label{CPW}
\int d \tau_0\,  \langle \mO_1\mO_2 \mO_{h}(\tau_0)\rangle \langle \mO_{1-h}(\tau_0) \mO_3 \mO_4\rangle =  \beta(h, h_{34}) \mF_{1234}^h (\tau_i) + \beta(1-h, h_{12}) \mF_{1234}^{1-h}(\tau_i)~, 
\ee
forming a conformal partial wave, which captures the exchange of $\mO_h$ and its descendants.~\footnote{$\mO_{1-h}$ is the shadow of $\mO_h$, and has dimensions $1-h$,  so  (\ref{CPW}) transforms as a four-point function.}  Explicitly evaluating the integral yields the right-hand side:  $\beta(h, h_{34})$ is a ratio of gamma functions, whose explicit form we have not written, and  $ \mF_{1234}^h (\tau_i)$ is identified as the conformal block, and contains a hypergeometric function of the conformally invariant cross ratio of the four times $\tau_1, \ldots, \tau_4$, and depends on the  dimensions of the external operators, $h_1, \ldots h_4$, and the dimension $h$ of the exchanged operator. 
  The conformal blocks form a basis, in terms of which one can express a general four-point function, 
\be \label{cBlock}
\langle \mO_1 \cdots \mO_4\rangle = \int_{\mC} \frac{d h}{2\pi i}\,  \rho (h) \mF_{1234}^h(\tau_i)~.
\ee
The contour $\mC$ runs parallel to the imaginary axis, $h = \frac{1}{2} + i s$, and in addition has counterclockwise circles enclosing the positive even integers $h = 2 n$. These are the principal and discrete series, respectively,  of $SL_2(R)$.~\footnote{For discussion of this in the context of SYK, see \cite{MS, PR, Kitaev:2017hnr}, as well as \cite{Murugan:2017eto, Bulycheva:2017uqj}. For a more general discussion, see  older work \cite{Dobrev:1977qv}, and more modern work \cite{Karateev:2017jgd, Caron-Huot:2017vep, Raben:2018sjl, Gadde:2017sjg}. For a discussion of conformal partial waves, see \cite{SimmonsDuffin:2012uy}. }

As an analogy, this expression for the four-point function is for the conformal group $SL_2(R)$ what the Fourier transform is for the translation group. Specifically, we may write any function of $x$ as, 
\be
f(x) = \int \frac{dp}{2\pi}\,  f(p) e^{i p x}~.
\ee
Here $e^{i p x}$ are a complete set of eigenfunctions of the Casimir of the translation group, $\partial_x^2$, while in (\ref{cBlock}), the conformal blocks $\mF_{1234}^h(\tau_i)$, with $h$ running over $\mC$,  are a complete set of eigenfunctions of the $SL_2(R)$ Casimir. 
Any CFT$_1$ four-point function is completely specified by an analytic function $\rho(h)$. The poles and residues of $\rho(h)$ set the dimensions and OPE coefficients of the exchanged operators in the four-point function, as one can see by closing the contour in (\ref{cBlock}).

For theories with $O(N)$ symmetry, it is natural to study operators that have definite transformation under the action of $O(N)$. We will be interested in $O(N)$ singlets, such as,~\footnote{This is schematic; some of the derivatives should act on the left $\chi_i$ as well as on the right $\chi_i$, in a specific way, so as to ensure the operator is primary. 
Also, we have not included operators with an even number of derivatives, since their correlation functions will vanish, by fermion antisymmetry. }
\be \label{ST}
\mO_{h}=\frac{1}{N} \sum_{i=1}^N \chi_i \partial_{\tau}^{1+2n} \chi_i~.
\ee
We will refer to such an operator as single-trace. One can make more $O(N)$ invariant operators, by taking products. For instance, a double-trace operator is schematically of the form $\mO_{h_1} \partial_{\tau}^{2n} \mO_{h_2}$.

\subsection{SYK correlation functions}

We discussed that the fermion two-point function is dominated by melons at large $N$. Now let us look at the Feynman  diagrams contributing to a connected fermion $2k$-point function. At leading order, these correlators will scale as $N^{-(k-1)}$, and will involve each external index occurring in pairs. Connecting two such lines by a propagator gives a Feynman diagram contributing to a $2(k\!-\!1)$-point function. Therefore, the Feynman diagrams for a $2k$-point function are found by successively cutting melon diagrams. 
\begin{figure}[t]
\centering 
\subfloat[]{
\includegraphics[width=1.6in]{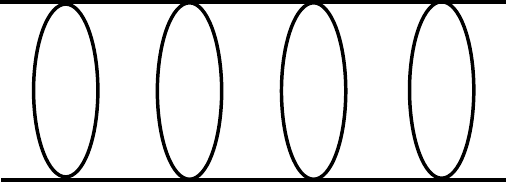}
}\hspace{2cm}
\subfloat[]{
\includegraphics[width=.53in]{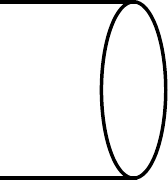}
}
\caption{(a) The four-point function is a sum of ladder diagrams. (b) The kernel that adds rungs to the ladder. 
Each line denotes the full propagator, so it is actually dressed by melons. }\label{Fig4pt}
\end{figure}

A single cut  gives the four-point function: it scales as $1/N$, and is a sum of ladder diagrams, shown in Fig.~\ref{Fig4pt}(a), with the kernel, shown in Fig.~\ref{Fig4pt}(b),  adding rungs to the ladder. To perform the sum, one need only find the eigenfunctions and eigenvalues of the kernel. In fact, as a result of $SL_2(R)$ invariance, the eigenfunctions are just the conformal partial waves discussed earlier, labeled by the dimension $h$ of the exchanged operator. In this basis, the sum becomes a geometric sum, and the fermion four-point function is of the form given in (\ref{cBlock}), with $\rho(h)$ \cite{MS}, \footnote{More precisely, the fermion four-point given here is defined as the $1/N$ piece of $N^{-2}\sum_{i,j =1}^N\langle \chi_i(\tau_1) \chi_i(\tau_2) \chi_j(\tau_3) \chi_j(\tau_4)\rangle$. A similar comment applies to higher point functions discussed later.}
\be \label{rho}
\rho(h)  = \mu(h) \frac{\alpha_0}{2} \frac{k(h)}{1 - k(h)}~, \ \ \ \ \  \ \ \ k(h) = - \frac{3}{2}\frac{\tan\frac{ \pi (h- 1/2)}{2}}{h - \frac{1}{2}}~,
\ee
where $k(h)$ are the eigenvalues of the kernel,  while $\mu(h)$ is a simple measure factor and $\alpha_0$ is a constant, which we have not written explicitly.~\footnote{See Eq.~2.17 and Eq.~2.26 of \cite{GR4} for the precise expression; in order to simplify, relative to the expression there, we have absorbed a factor of $\Gamma(h)^2/\Gamma(2h)$ into the definition of $\mu(h)$, and neglected a factor of $b^2/J^{4\Delta}$. Also,  equation (\ref{4ptO}) on the next page is Eq.~4.16 of \cite{GR4}.  }  As mentioned earlier, the poles and residues of $\rho(h)$ give the dimensions and OPE coefficients, $c_h$,  of the exchanged operators. The poles  occur at the $h$ for which $k(h)=1$.~\footnote{The integral equation determining $k(h)$ is like a Bethe-Salpeter equation for conformal theories; instead of the masses of the bound states, it determines the dimensions $h$ of the composite operators. See \cite{GR}.  In some places in the literature, $k(h)$ is instead denoted by $k_c(h)$, or  by $g(h)$.}
These can be written in the form, $h = 2 n+1 + 2\Delta + 2\eps_n$, with $\eps_n$ small for large integer $n$, and correspond to the dimensions of the single-trace operators (\ref{ST}) (in the infrared).~\footnote{The location of the smallest positive $h$ for which $k(h)=1$ is $h=2$. The $h=2$ operator is special: it lies on the contour $\mC$, and so leads to a divergence. This means we must move slightly away from the infinite $J$ limit. For cSYK \cite{GR3}, which is conformal at any value of $J$, the four-point function at large but finite $J$ is given by an expression similar to (\ref{rho}), but for which $k(h)=1$ at an $h$ slightly less than two. This makes the  contribution of this block finite but large. In SYK, moving to large but finite $J$ means breaking conformal invariance and accounting for the Schwarzian action. The contribution of the $h=2$ block will come with a factor of $J$, relative to the other blocks, and so it dominates the four-point function. The Lyaponuv exponent from the $h=2$ contribution is maximal \cite{MSS}, and since this piece dominates, SYK is maximally chaotic at large $J$ \cite{Kitaev}. \label{4ptFoot} }
\begin{figure}[t]
\centering
\subfloat[]{
\includegraphics[width=3.8in]{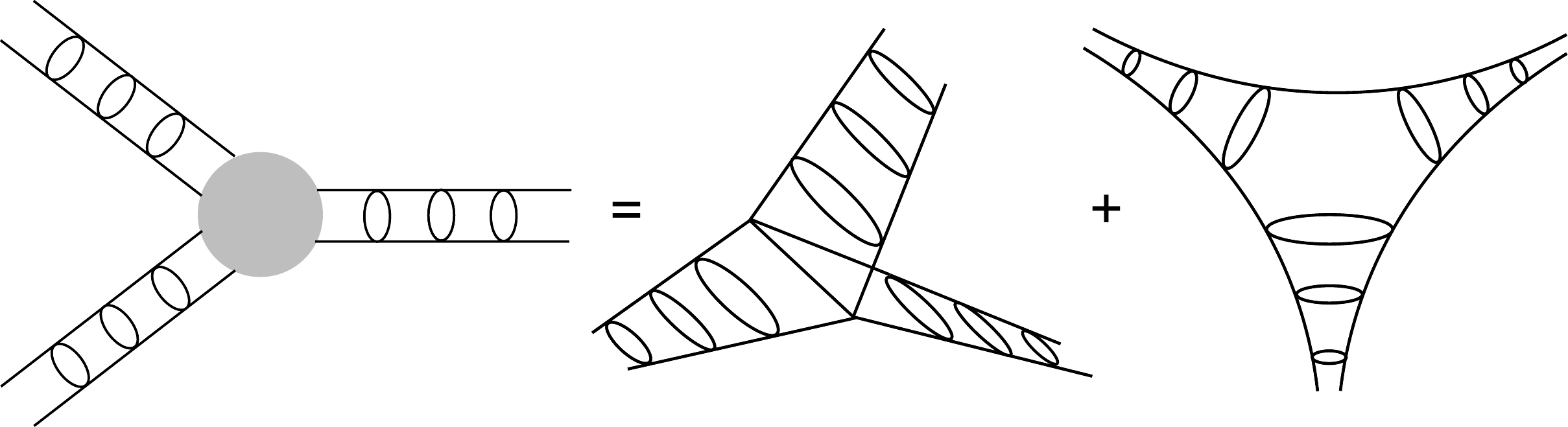}
} \ \ \ \ \ \   \ \ 
\subfloat[]{
\includegraphics[width=2in]{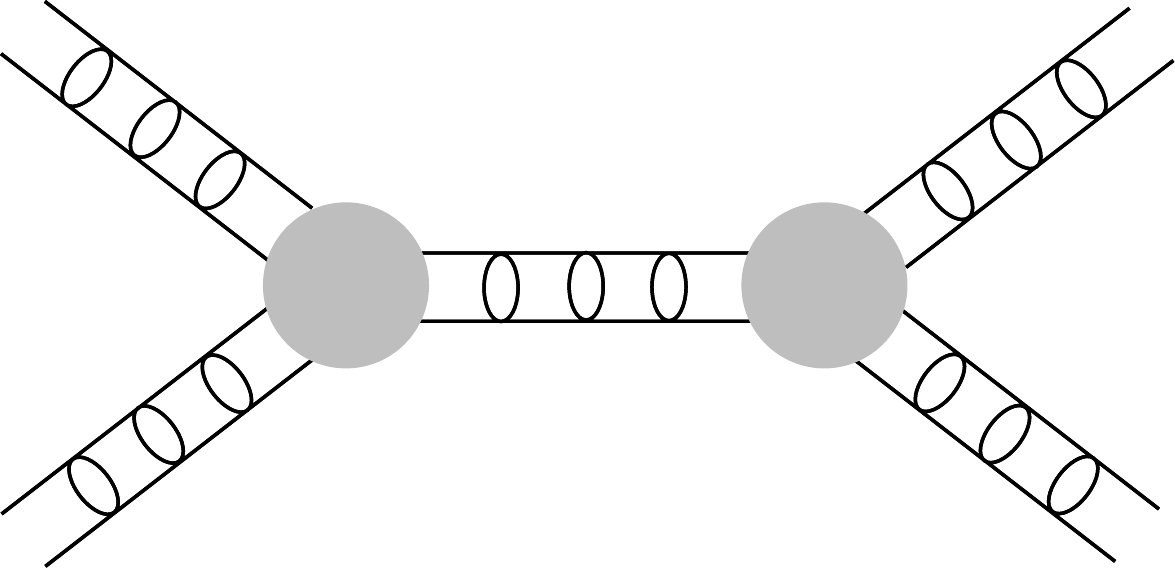}
}
\caption{(a) The fermion six-point function. (b) A contribution to the eight-point function.  } \label{Fig68}
\end{figure}

The fermion six-point function consists of the diagrams shown in Fig.~\ref{Fig68}(a), and can be viewed as three four-point functions  glued together. Since the four-point function is expressed in terms of conformal blocks, computing the six-point function is just a matter of gluing together three conformal blocks. The information in the fermion six-point function can be compactly encoded in the three-point functions of  the bilinear operators $\mO_h$ (\ref{ST}), with coefficients $C_{h_1 h_2 h_3}$, whose explicit form is given in \cite{GR4}: they  can be written as $C_{h_1 h_2 h_3} = c_{h_1} c_{h_2} c_{h_3} \mI_{h_1 h_2 h_3}$, where $\mI_{h_1 h_2 h_3}$  is an analytic function of the $h_i$, involving gamma functions and the hypergeometric function ${}_4 F_3$ at argument one. 

The fermion eight-point function  is built out of more four-point functions glued together. One such contribution is shown in Fig.~\ref{Fig68}(b), and its contribution to the bilinear four-point function takes an incredibly simple form \cite{GR4},
\be \label{4ptO}
c_{h_1} c_{h_2} c_{h_3} c_{h_4} \int_{\mC} \frac{d h}{2\pi i} \, \rho (h)\,  \mI_{h_1 h_2\, h}\, \mI_{h\, h_3 h_4}\, \mF_{1234}^h(\tau_i)~.
\ee
The result is intuitive: the $\mI_{h_1 h_2\, h}$ and $\mI_{h\,  h_3 h_4}$ are the ``interaction vertices'' from the two six-point functions, and there is an operator of dimension $h$ exchanged, giving the $\rho(h)$ factor from the intermediate fermion four-point function. A nontrivial consistency check is that the four-point function of $\mO_{h_i}$, at order $1/N$, should be a sum of conformal blocks of exchanged single-trace operators as well as double-trace operators. Upon closing the contour in (\ref{4ptO}), the poles of $\rho(h)$ give the single-trace blocks, while the poles of $\mI_{h_1 h_2 h}$, occurring at  $h= h_1 + h_2 +2n$, give the double-trace blocks. It is remarkable that the analytically extended OPE coefficients of the single-trace operators - the $\mI_{h_1 h_2 h_3}$ - knew  that they should have singularities at precisely these locations. 

While SYK has a special set of Feynman diagrams, these results for the correlation functions are more general. The simple expression for the fermion four-point function follows from summing ladder diagrams; it is irrelevant that the propagators are built from melons, those only served to give a conformal two-point function.~\footnote{In fact, similar ladder diagrams appear in the fishnet theory, a deformation of $\mathcal{N}=4$ super Yang-Mills \cite{Grabner:2017pgm}.} The six-point point function is made up of three four-point functions glued together, and in calculating it, it is not relevant that the four-point function was a sum of ladder diagrams. The expression for the eight-point function/bilinear four-point function is valid regardless of the details of how the three fermion four-point functions combine at the ``interaction vertex''; all of this is encoded in $\mI_{h_1 h_2 h_3}$.

\section{Applications}\label{sec:applications}

\subsection{AdS/CFT} \label{adscft}

At low energies, SYK is dominated by the $h=2$ mode, described by the Schwarzian action. This is a result of being nearly conformally invariant. 
On the AdS$_2$ side, since Einstein gravity is topological in two-dimensions, it is natural to instead consider Jackiw-Teitelboim dilaton gravity  \cite{Jackiw:1984je, AP}. Dilaton gravity theories naturally arise from compactifying gravity in higher dimensions down to two dimensions, with the dilaton playing the role of the size of the extra dimension.  It has been shown that the dilaton theory in AdS$_2$ is the same as the Schwarzian theory, as a consequence of the pattern of symmetry breaking \cite{MSY, Jensen:2016pah, Engelsoy:2016xyb}.~\footnote{For further studies of two-dimensional gravity, AdS$_2$, and the Schwarzian, see\cite{ Dubovsky:2017cnj, Forste:2017apw,  Haehl:2017pak, Grumiller:2017qao,Gonzalez:2018enk,  Lam:2018pvp,Qi:2018rqm,  Taylor:2017dly, Nayak:2018qej, Li:2018omr}. }

Of course, the $h=2$ mode is just the first in the tower of fermion bilinear $O(N)$ singlets, $\mO_{h}$, written schematically in (\ref{ST}), and the rest of the tower, with $n\geq 1$, encode the structure of SYK.  
As we have discussed,  a connected $k$-point correlation function of the $\mO_h$ scales as $N^{-(k-2)/2}$. 
We can write a putative dual field theory in AdS$_2$, 
\be \label{Lbulk}
\mathcal{L}_{bulk} =  \sum_{n=1}^{\infty} \frac{1}{2} (\partial \phi_n)^2 + \frac{1}{2} m_n^2 \phi_n^2 + \frac{1}{\sqrt{N}} \sum_{n, m, k=1}^{\infty}\lambda_{n m k } \phi_n \phi_m \phi_k+ \ldots
\ee
containing a tower of scalar fields $\phi_n$. As a result of the $SL_2(R)$ isometry of AdS$_2$, any correlation function of the $\phi_n$, at points extrapolated to  the boundary of AdS$_2$, will take the form of a CFT correlation function. Identifying each $\phi_n$ with  an operator $\mO_h$, we can appropriately choose the masses and cubic couplings of the $\phi_n$ so as to match to the SYK two-point and three-point functions of the $\mO_{h}$, respectively. The masses are related to the dimensions in the standard way, $m_n^2 = h(h- 1)$, while the cubic couplings,  in the limit $n, m, k \gg 1$ where they simplify, are \cite{GR4, GR2},
\be \label{lambda}
\lambda_{n m k} \approx \frac{(n + m +k)!}{\Gamma(n+m-k+\frac{1}{2}) \Gamma(m+k-n+\frac{1}{2})\Gamma(k+n -m+ \frac{1}{2})}~.
\ee 
\begin{figure}[t]
\centering
\includegraphics[width=5in]{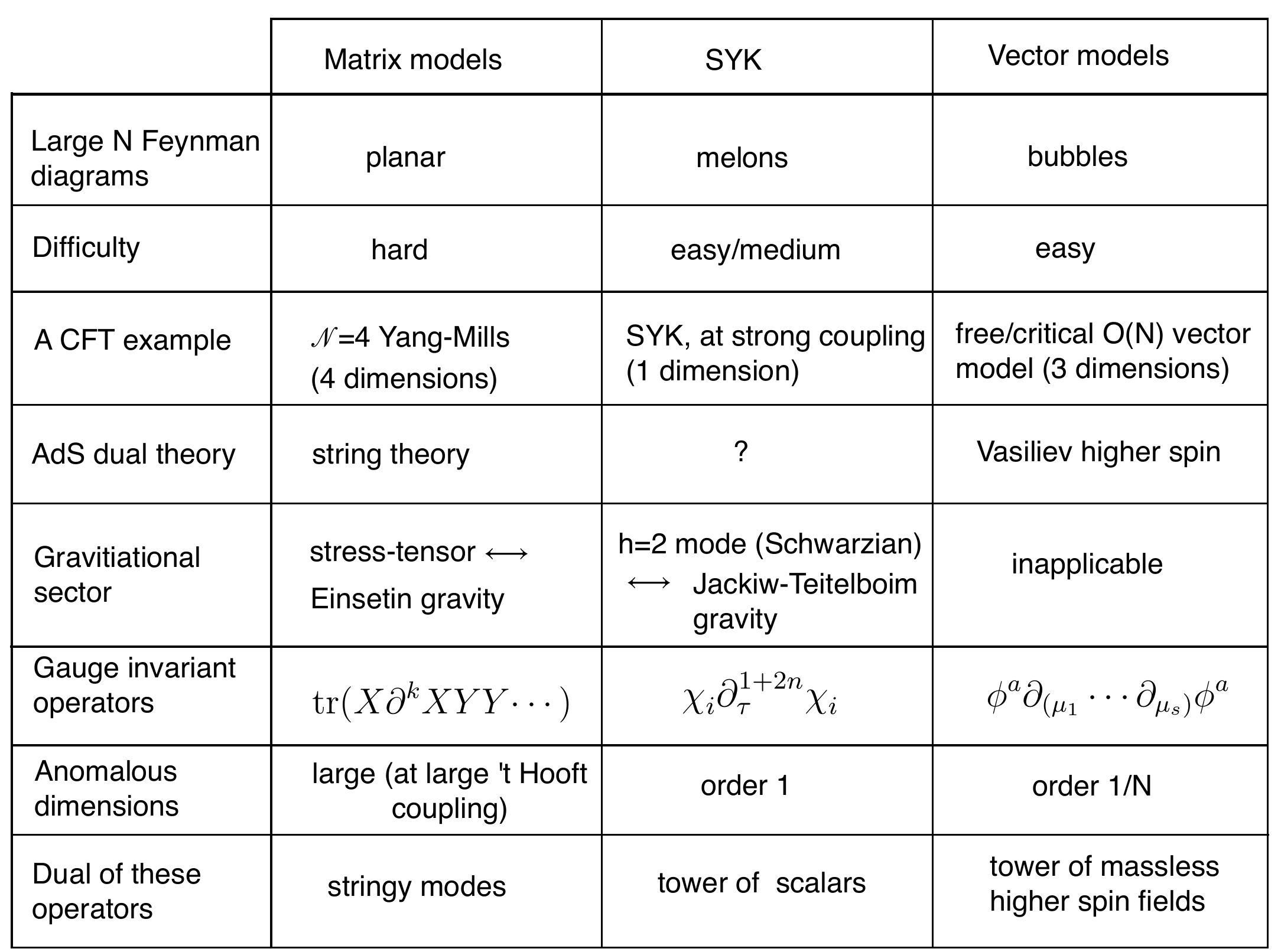}
\caption[]{A comparison of SYK to matrix models and vector models. One comment is that for $\mathcal{N}=4$ the anomalous dimensions are large only for the non-supersymmetry protected operators (the half-BPS operators are dual to Kaluza-Klein modes on the $S^5$; such operators are ignored in the table). Another, relating to the gravity description,  is that it is only at large 't Hooft coupling, when the stringy modes become very massive, that one can say that the bulk is Einstein gravity. This is to be contrasted with the bulk dual of SYK, where there is no limit in which the tower of bulk fields decouple; their mass is of order-one. Finally,  in Vasiliev theory, the spin two field (graviton) is related by symmetry to the other higher spin fields. } \label{FigTable}
\end{figure}
\indent One could similarly try to appropriately choose the quartic couplings, so as to match the SYK four-point function of bilinears.~\footnote{The leading connected SYK correlators that we computed in the previous section map onto the tree-level Witten diagrams. The $1/\sqrt{N}$ corrections to these would map onto loops in the bulk, and so are not needed in order to establish the classical bulk Lagrangian (\ref{Lbulk}). } However, in order to have an actual understanding of the AdS dual of SYK, one needs a simple and independently defined bulk theory, something like the string worldsheet action, rather than just a list of couplings. Such a bulk description is presently lacking;  it is not  obvious one must  exist.~\footnote{Some proposals are as follows: A single scalar field  Kaluza-Klein reduced on an AdS$_2 \times S^1$, or something like it, can be made to give the correct spectrum of masses but gives the wrong cubic couplings \cite{GR4, Das:2017hrt}. A string with longitudinal motion \cite{Bardeen:1975gx, MS} also gives a qualitatively correct spectrum, but such solutions are only known at the classical level; one would need a quantum theory, in order to determine the cubic couplings. The 't Hooft model of two-dimensional QCD, but placed in AdS, is  perhaps the most promising, but is difficult to solve \cite{GR5}, and would at best match SYK only qualitatively, with no a priori reason it should match exactly. 
One might instead search for the bulk dual of the tensor models, rather than SYK, but tensor models have a vast number of singlets, and the bulk dual would correspondingly  have a huge number of  fields and a Hagedorn temperature scaling as $1/\log N$ \cite{Bulycheva:2017ilt, Choudhury:2017tax, Beccaria:2017aqc}, see also \cite{Krishnan:2017lra}, and the bulk description would  likely be even more complicated than Vasiliev theory \cite{Vasiliev:2018zer}.  }

Two canonical examples of AdS/CFT duality are between  $\mathcal{N} =4$ super Yang-Mills in 4 dimensions and string theory in AdS$_5 \times S^5$ \cite{Maldacena:1997re, Gubser:1998bc, Witten:1998qj}, and between the free/critical vector $O(N)$ model in 3 dimensions and Vasiliev higher spin theory in AdS$_4$ \cite{Klebanov:2002ja, Giombi:2016ejx}. A comparison between SYK and these two theories is given in Table \ref{FigTable}.

One hope for SYK has been that, because of its simplicity,  it would provide an example of AdS/CFT  in which one could fully understand the duality, directly relating the CFT degrees of freedom to the bulk variables.  This remains a goal, though achieving it of course requires knowing what the bulk theory is, in order to have a target. Independently of this, SYK has led  to a renewed interest in two-dimensional gravity, and the formulation of modern ideas on spacetime and holography in this context, see e.g. \cite{Ooguri:2016pdq, Cotler:2016fpe, Saad:2018bqo, Maldacena:2017axo, Kourkoulou:2017zaj ,Maldacena:2018lmt,Brown:2017jil, Caputa:2017yrh, Harlow:2018tqv, Swingle}.

\subsection{Strange metals}
There are  many variants of SYK, which retain the key feature of dominance of melon diagrams. 
One natural generalization, which incorporates some of these, is to consider a model  which contains $f$ flavors of fermions, with $N_a$ fermions of flavor $a$, each appearing $q_a$ times in the interaction, so that the Hamiltonian couples $\mathrm{q}=\sum_{a=1}^f q_a$ fermions together \cite{GR},
\be \label{GFM}
L= \frac{1}{2}\sum_{a =1}^f \sum_{i=1}^{N_a}\chi_i^a\, \partial_{\tau}\chi_i^a +\frac{(i)^{\frac{\mathrm{q}}{2}}}{\prod_{a=1}^f q_a!} \sum_I J_{I}( \chi_{i_1}^1 \cdots \chi_{i_{q_1}}^1)\cdots (\chi_{j_1}^f \cdots \chi_{j_{q_f}}^f) ~,
\ee
where $I$ is a collective index, $I = i_1,\ldots, i_{q_1}, \ldots, j_{1},\ldots, j_{q_f}$. The coupling $J_{I}$ is antisymmetric under permutation of indices within any one of the $f$ families, and  is drawn from a Gaussian distribution. This model with one flavor, $f=1$, reduces to the standard SYK model with a $q$-body interaction, sometimes denoted by SYK$_q$; further setting $q=4$ gives the canonical SYK model, which has been the focus of these notes.~\footnote{Starting with the flavored model, and taking $f=2$ and $q_1=1$, and replacing  the first fermion with an auxiliary boson, gives the supersymmetric SYK model \cite{Fu:2016vas}.  See \cite{Sannomiya:2016mnj, Peng:2017spg,  Bulycheva:2018qcp, Narayan:2017hvh} for further studies, and  \cite{Anninos:2016szt} for an earlier string inspired model. Supersymmetry has so far been important in the construction of SYK models in higher dimensions  \cite{Murugan:2017eto}, unless perhaps one works in non-integer dimension \cite{Giombi:2017dtl, Klebanov:2016xxf}, see also \cite{Prakash:2017hwq, Benedetti:2017fmp, Azeyanagi:2017drg}. Constructing a conformal tensor version of the simplest supersymmetric SYK model, with interaction $J_{i j k} \phi_i \chi_j \chi_k$,  is nontrivial \cite{Peng:2016mxj, Chang:2018sve}.  Some other variations of SYK include perturbed SYK models \cite{Bi:2017yvx}, chiral models \cite{Berkooz:2016cvq,Peng:2018zap}, models with non-standard kinetic terms \cite{Turiaci:2017zwd},  and $p$-adic models \cite{Gubser:2017qed}.}

\begin{figure}[t]
\vspace{-.5cm}
\centering
\includegraphics[width=1.7in]{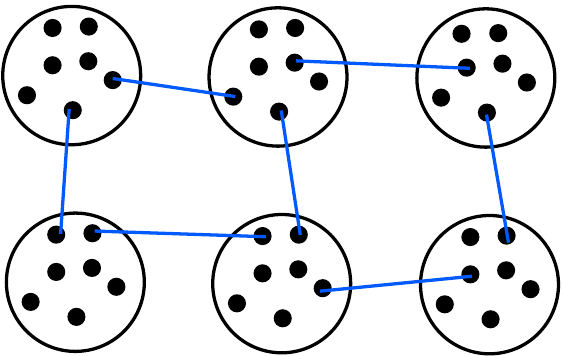}
\caption{A lattice of SYK ``quantum dots'', with quartic all-to-all interactions within a dot, the third term in (\ref{dot}), and quadratic hopping terms between dots, the second term in (\ref{dot}).}\label{FigLattice}
\vspace{-.2cm}
\end{figure}

Regarding flavor as a lattice site index, $x$, and taking sums of the flavored models, one can build lattices of SYK models \cite{Gu:2016oyy}.   One such model, having some features of a strongly correlated metal, is \cite{Song:2017pfw},\vspace{-.2cm}
\be \label{dot}
L = \sum_{x}\sum_i c_{i, x}^{\dagger} \(\partial_{\tau} - \mu\) c_{i, x} - \sum_{\langle x x'\rangle} \sum_{i, j} t_{ij, x x'} c_{i, x}^{\dagger} c_{j, x'} - \sum_{x} \sum_{i, j, k, l} J_{i j k l, x}c_{i, x}^{\dagger} c_{j, x}^{\dagger} c_{k, x} c_{l, x}~.
\vspace{-.2cm}
\ee 
Here $t_{ij, x x'}$ and $J_{i j k l, x}$ are again random couplings,  and the $c_{i, x}$ are now complex fermions. A cartoon of this model is shown in Fig.~\ref{FigLattice}. The model exhibits incoherent metal behavior, with resistivity scaling linearly with temperature, at high temperature, and Fermi liquid behavior at low temperature.~\footnote{ For other studies of SYK lattice models, see \cite{Banerjee:2016ncu, Jian:2017unn, Chowdhury:2018sho,Fu:2018spl, Patel:2018rjp}, and for a lattice model exploiting tensors, see \cite{Wu:2018vua}.}

More generally, the fact that SYK is a system without quasiparticles, yet is nevertheless solvable, makes it a valuable tool with which to study  transport and chaos \cite{Davison:2016ngz,Gu:2017ohj, Peng:2017kro, Chen:2017dbb, Patel:2017mjv, Mondal:2018xwy, Blake:2016jnn,   Kukuljan:2017xag, Scaffidi:2017ghs}, non-equilibrium dynamics and entanglement  \cite{Liu:2017kfa, Gu:2017njx, Eberlein:2017wah, Huang:2017nox}, and eigenstate thermalization \cite{Sonner:2017hxc, Haque:2017bts, Hunter-Jones:2017raw}.  There are limitations, however, as neither the all-to-all interactions nor the large $N$, which are essential to the solvability of SYK, are present in  real metals.

\vspace{.3cm}
Finally, there are a number of topics which we have not discussed, such as: experimental realizations and quantum simulations of SYK \cite{Danshita:2016xbo, Garcia-Alvarez:2016wem,   Pikulin:2017mhj, Chew:2017xuo, Luo:2017bno, Chen:2018wvb, 2018arXiv180602793B},  the zero-temperature entropy (an infinite $N$ artifact)  \cite{Parcollet:1997ysb, Sachdev:2010um, Sachdev:2015efa}, studies of the spectral density, the spectral form factor, and connections with random matrix theory \cite{Cotler:2016fpe, Garcia-Garcia:2016mno, You:2016ldz,  Krishnan:2016bvg, Liu:2016rdi, Li:2017hdt, Chaudhuri:2017vrv, Kanazawa:2017dpd,  Garcia-Garcia:2017bkg, Kos:2017zjh, Altland:2017eao, Feng:2018zsx }.

\vspace{-.31cm}
\section*{Acknowledgements} 
\vspace{-.2cm}

\noindent I am very grateful to David Gross for collaboration. 
I  thank  A.~Kitaev, I.~Klebanov, J.~Maldacena,   S.~Sachdev, J.~Suh, and H.~Verlinde for many discussions. 
I am grateful for having had the opportunity to collaborate with Joe Polchinski, an extraordinary theorist who had a profound impact on physics, and on me personally. 
This work was supported by NSF grant 1125915.

\begin{changemargin}{-1cm}{-.75cm}
\patchcmd{\thebibliography}
  {\settowidth}
  {\setlength{\itemsep}{0pt plus 0.1pt}\settowidth}
  {}{}
\apptocmd{\thebibliography}
  {\footnotesize}
  {}{}

\bibliographystyle{utphys}

{\setstretch{.935}


\begin{thebibliography}{100}

\bibitem{SY}
S.~Sachdev and J.~Ye, ``Gapless spin-fluid ground state in a random quantum
  heisenberg magnet,''
  \href{http://dx.doi.org/10.1103/PhysRevLett.70.3339}{{\em Phys. Rev. Lett.}
  {\bfseries 70} (May, 1993) 3339--3342}.
  \url{http://link.aps.org/doi/10.1103/PhysRevLett.70.3339}.

\bibitem{Kitaev}
A.~Kitaev, ``{A simple model of quantum holography},'' {\em KITP strings
  seminar and Entanglement 2015 program} (Feb. 12, April 7, and May 27, 2015) .
  \url{http://online.kitp.ucsb.edu/online/entangled15/}.

\bibitem{Coleman}
S.~Coleman, {\em {Aspects of Symmetry}}.
\newblock Cambridge University Press, 1985.

\bibitem{Gross:1974jv}
D.~J. Gross and A.~Neveu, ``{Dynamical Symmetry Breaking in Asymptotically Free
  Field Theories},''
\href{http://dx.doi.org/10.1103/PhysRevD.10.3235}{{\em Phys. Rev.} {\bfseries
  D10} (1974) 3235}.

\bibitem{Iizuka:2008hg}
N.~Iizuka and J.~Polchinski, ``{A Matrix Model for Black Hole
  Thermalization},''
  \href{http://dx.doi.org/10.1088/1126-6708/2008/10/028}{{\em JHEP} {\bfseries
  10} (2008) 028},
\href{http://arxiv.org/abs/0801.3657}{{\ttfamily arXiv:0801.3657 [hep-th]}}.

\bibitem{tHooft:1974pnl}
G.~'t~Hooft, ``{A Two-Dimensional Model for Mesons},''
\href{http://dx.doi.org/10.1016/0550-3213(74)90088-1}{{\em Nucl. Phys.}
  {\bfseries B75} (1974) 461--470}.

\bibitem{Callan:1975ps}
C.~G. Callan, Jr., N.~Coote, and D.~J. Gross, ``{Two-Dimensional Yang-Mills
  Theory: A Model of Quark Confinement},''
\href{http://dx.doi.org/10.1103/PhysRevD.13.1649}{{\em Phys. Rev.} {\bfseries
  D13} (1976) 1649}.

\bibitem{Giombi:2011kc}
S.~Giombi, S.~Minwalla, S.~Prakash, S.~P. Trivedi, S.~R. Wadia, and X.~Yin,
  ``{Chern-Simons Theory with Vector Fermion Matter},''
  \href{http://dx.doi.org/10.1140/epjc/s10052-012-2112-0}{{\em Eur. Phys. J.}
  {\bfseries C72} (2012) 2112},
\href{http://arxiv.org/abs/1110.4386}{{\ttfamily arXiv:1110.4386 [hep-th]}}.

\bibitem{Aharony:2011jz}
O.~Aharony, G.~Gur-Ari, and R.~Yacoby, ``{d=3 Bosonic Vector Models Coupled to
  Chern-Simons Gauge Theories},''
  \href{http://dx.doi.org/10.1007/JHEP03(2012)037}{{\em JHEP} {\bfseries 03}
  (2012) 037},
\href{http://arxiv.org/abs/1110.4382}{{\ttfamily arXiv:1110.4382 [hep-th]}}.

\bibitem{tHooft:1973alw}
G.~'t~Hooft, ``{A Planar Diagram Theory for Strong Interactions},''
  \href{http://dx.doi.org/10.1016/0550-3213(74)90154-0}{{\em Nucl. Phys.}
  {\bfseries B72} (1974) 461}.
[,337(1973)].

\bibitem{DiFrancesco:1993cyw}
P.~Di~Francesco, P.~H. Ginsparg, and J.~Zinn-Justin, ``{2-D Gravity and random
  matrices},'' 
\href{http://arxiv.org/abs/hep-th/9306153}{{\ttfamily arXiv:hep-th/9306153}}

\bibitem{Klebanov:1991qa}
I.~R. Klebanov, ``{String theory in two-dimensions},'' \href{http://arxiv.org/abs/hep-th/9108019}{{\ttfamily arXiv:hep-th/9108019
  [hep-th]}}.
\newblock

\bibitem{Beisert:2010jr}
N.~Beisert {\em et~al.}, ``{Review of AdS/CFT Integrability: An Overview},''
\href{http://arxiv.org/abs/1012.3982}{{\ttfamily arXiv:1012.3982 [hep-th]}}.

\bibitem{Gromov:2013pga}
N.~Gromov, V.~Kazakov, S.~Leurent, and D.~Volin, ``{Quantum Spectral Curve for
  Planar $\mathcal{N} = 4$ Super-Yang-Mills Theory},''
  \href{http://dx.doi.org/10.1103/PhysRevLett.112.011602}{{\em Phys. Rev.
  Lett.} {\bfseries 112} no.~1, (2014) 011602},
\href{http://arxiv.org/abs/1305.1939}{{\ttfamily arXiv:1305.1939 [hep-th]}}.

\bibitem{Basso:2015zoa}
B.~Basso, S.~Komatsu, and P.~Vieira, ``{Structure Constants and Integrable
  Bootstrap in Planar N=4 SYM Theory},''
\href{http://arxiv.org/abs/1505.06745}{{\ttfamily arXiv:1505.06745 [hep-th]}}.

\bibitem{Zamolodchikov:1978xm}
A.~B. Zamolodchikov and A.~B. Zamolodchikov, ``{Factorized s Matrices in
  Two-Dimensions as the Exact Solutions of Certain Relativistic Quantum Field
  Models},'' \href{http://dx.doi.org/10.1016/0003-4916(79)90391-9}{{\em Annals
  Phys.} {\bfseries 120} (1979) 253--291}.

\bibitem{Minahan:2002ve}
J.~A. Minahan and K.~Zarembo, ``{The Bethe ansatz for N=4 superYang-Mills},''
  \href{http://dx.doi.org/10.1088/1126-6708/2003/03/013}{{\em JHEP} {\bfseries
  03} (2003) 013},
\href{http://arxiv.org/abs/hep-th/0212208}{{\ttfamily arXiv:hep-th/0212208
  [hep-th]}}.

\bibitem{Klebanov:2016xxf}
I.~R. Klebanov and G.~Tarnopolsky, ``{Uncolored random tensors, melon diagrams,
  and the Sachdev-Ye-Kitaev models},''
  \href{http://dx.doi.org/10.1103/PhysRevD.95.046004}{{\em Phys. Rev.}
  {\bfseries D95} no.~4, (2017) 046004},
\href{http://arxiv.org/abs/1611.08915}{{\ttfamily arXiv:1611.08915 [hep-th]}}.

\bibitem{Gurau:2009tw}
R.~Gurau, ``{Colored Group Field Theory},''
  \href{http://dx.doi.org/10.1007/s00220-011-1226-9}{{\em Commun. Math. Phys.}
  {\bfseries 304} (2011) 69--93},
\href{http://arxiv.org/abs/0907.2582}{{\ttfamily arXiv:0907.2582 [hep-th]}}.

\bibitem{Witten:2016iux}
E.~Witten, ``{An SYK-Like Model Without Disorder},''
\href{http://arxiv.org/abs/1610.09758}{{\ttfamily arXiv:1610.09758 [hep-th]}}.

\bibitem{Bonzom:2011zz}
V.~Bonzom, R.~Gurau, A.~Riello, and V.~Rivasseau, ``{Critical behavior of
  colored tensor models in the large N limit},''
  \href{http://dx.doi.org/10.1016/j.nuclphysb.2011.07.022}{{\em Nucl. Phys.}
  {\bfseries B853} (2011) 174--195},
\href{http://arxiv.org/abs/1105.3122}{{\ttfamily arXiv:1105.3122 [hep-th]}}.

\bibitem{2016LMaPh.106.1531C}
S.~{Carrozza} and A.~{Tanasa}, ``{O( N) Random Tensor Models},''
  \href{http://dx.doi.org/10.1007/s11005-016-0879-x}{{\em Letters in
  Mathematical Physics} {\bfseries 106} (Nov., 2016) 1531--1559},
  \href{http://arxiv.org/abs/1512.06718}{{\ttfamily arXiv:1512.06718
  [math-ph]}}.
  

  
\bibitem{Delporte:2018iyf}
N.~Delporte and V.~Rivasseau, ``{The Tensor Track V: Holographic Tensors},''
\href{http://arxiv.org/abs/1804.11101}{{\ttfamily arXiv:1804.11101 [hep-th]}}.

\bibitem{Klebanov:2017nlk}
I.~R. Klebanov and G.~Tarnopolsky, ``{On Large $N$ Limit of Symmetric Traceless
  Tensor Models},'' \href{http://dx.doi.org/10.1007/JHEP10(2017)037}{{\em JHEP}
  {\bfseries 10} (2017) 037},
\href{http://arxiv.org/abs/1706.00839}{{\ttfamily arXiv:1706.00839 [hep-th]}}.

  
\bibitem{Gurau:2017xhf}
R.~Gurau, ``{Quenched equals annealed at leading order in the colored SYK
  model},'' \href{http://dx.doi.org/10.1209/0295-5075/119/30003}{{\em EPL}
  {\bfseries 119} no.~3, (2017) 30003},
\href{http://arxiv.org/abs/1702.04228}{{\ttfamily arXiv:1702.04228 [hep-th]}}.

\bibitem{MPRS}
B.~Michel, J.~Polchinski, V.~Rosenhaus, and S.~J. Suh, ``{Four-point function
  in the IOP matrix model},''
  \href{http://dx.doi.org/10.1007/JHEP05(2016)048}{{\em JHEP} {\bfseries 05}
  (2016) 048},
\href{http://arxiv.org/abs/1602.06422}{{\ttfamily arXiv:1602.06422 [hep-th]}}.

\bibitem{MS}
J.~Maldacena and D.~Stanford, ``{Comments on the Sachdev-Ye-Kitaev model},''
\href{http://arxiv.org/abs/1604.07818}{{\ttfamily arXiv:1604.07818 [hep-th]}}.


\bibitem{GR}
D.~J. Gross and V.~Rosenhaus, ``{A Generalization of Sachdev-Ye-Kitaev},''
  \href{http://dx.doi.org/10.1007/JHEP02(2017)093}{{\em JHEP} {\bfseries 02}
  (2017) 093},
\href{http://arxiv.org/abs/1610.01569}{{\ttfamily arXiv:1610.01569 [hep-th]}}.

\bibitem{Kitaev:2017awl}
A.~Kitaev and S.~J. Suh, ``{The soft mode in the Sachdev-Ye-Kitaev model and
  its gravity dual},''
\href{http://arxiv.org/abs/1711.08467}{{\ttfamily arXiv:1711.08467 [hep-th]}}.

\bibitem{DMFT}
A.~Georges, G.~Kotliar, W.~Krauth, and M.~J. Rozenberg, 
   \href{http://dx.doi.org/10.1103/RevModPhys.68.13}{{\em Rev.
  Mod. Phys.} {\bfseries 68}
  (Jan, 1996) 13--125 }
; A.~Georges, ``{Applications of Dynamical Mean Field Theory},'' 
  (Aug. 20, 2015) .
  \url{http://online.kitp.ucsb.edu/online/latticeqcd15/dmft/}.

\bibitem{PR}
J.~Polchinski and V.~Rosenhaus, ``{The Spectrum in the Sachdev-Ye-Kitaev
  Model},'' \href{http://dx.doi.org/10.1007/JHEP04(2016)001}{{\em JHEP}
  {\bfseries 04} (2016) 001},
\href{http://arxiv.org/abs/1601.06768}{{\ttfamily arXiv:1601.06768 [hep-th]}}.

\bibitem{MSY}
J.~Maldacena, D.~Stanford, and Z.~Yang, ``{Conformal symmetry and its breaking
  in two dimensional Nearly Anti-de-Sitter space},''
\href{http://arxiv.org/abs/1606.01857}{{\ttfamily arXiv:1606.01857 [hep-th]}}.


\bibitem{Jevicki:2016bwu}
A.~Jevicki, K.~Suzuki, and J.~Yoon, ``{Bi-Local Holography in the SYK Model},''
  \href{http://dx.doi.org/10.1007/JHEP07(2016)007}{{\em JHEP} {\bfseries 07}
  (2016) 007},
\href{http://arxiv.org/abs/1603.06246}{{\ttfamily arXiv:1603.06246 [hep-th]}}.; 
A.~Jevicki and K.~Suzuki, ``{Bi-Local Holography in the SYK Model:
  Perturbations},'' \href{http://dx.doi.org/10.1007/JHEP11(2016)046}{{\em JHEP}
  {\bfseries 11} (2016) 046},
\href{http://arxiv.org/abs/1608.07567}{{\ttfamily arXiv:1608.07567 [hep-th]}}.

\bibitem{Bagrets:2016cdf}
D.~Bagrets, A.~Altland, and A.~Kamenev, ``{Sachdev-Ye-Kitaev model as Liouville
  quantum mechanics},''
  \href{http://dx.doi.org/10.1016/j.nuclphysb.2016.08.002}{{\em Nucl. Phys.}
  {\bfseries B911} (2016) 191--205},
\href{http://arxiv.org/abs/1607.00694}{{\ttfamily arXiv:1607.00694
  [cond-mat.str-el]}};  
D.~Bagrets, A.~Altland, and A.~Kamenev, ``{Power-law out of time order
  correlation functions in the SYK model},''
  \href{http://dx.doi.org/10.1016/j.nuclphysb.2017.06.012}{{\em Nucl. Phys.}
  {\bfseries B921} (2017) 727--752},
\href{http://arxiv.org/abs/1702.08902}{{\ttfamily arXiv:1702.08902
  [cond-mat.str-el]}}.



\bibitem{Stanford:2017thb}
D.~Stanford and E.~Witten, ``{Fermionic Localization of the Schwarzian
  Theory},'' \href{http://dx.doi.org/10.1007/JHEP10(2017)008}{{\em JHEP}
  {\bfseries 10} (2017) 008},
\href{http://arxiv.org/abs/1703.04612}{{\ttfamily arXiv:1703.04612 [hep-th]}}.

\bibitem{Mertens:2017mtv}
T.~G. Mertens, G.~J. Turiaci, and H.~L. Verlinde, ``{Solving the Schwarzian via
  the Conformal Bootstrap},''
  \href{http://dx.doi.org/10.1007/JHEP08(2017)136}{{\em JHEP} {\bfseries 08}
  (2017) 136},
\href{http://arxiv.org/abs/1705.08408}{{\ttfamily arXiv:1705.08408 [hep-th]}}.

\bibitem{Mertens:2018fds}
T.~G. Mertens, ``{The Schwarzian Theory - Origins},''
\href{http://arxiv.org/abs/1801.09605}{{\ttfamily arXiv:1801.09605 [hep-th]}}; 
A.~Blommaert, T.~G. Mertens, and H.~Verschelde, ``{The Schwarzian Theory - A
  Wilson Line Perspective},''
\href{http://arxiv.org/abs/1806.07765}{{\ttfamily arXiv:1806.07765 [hep-th]}}.



\bibitem{Mandal:2017thl}
G.~Mandal, P.~Nayak, and S.~R. Wadia, ``{Coadjoint orbit action of Virasoro
  group and two-dimensional quantum gravity dual to SYK/tensor models},''
 \href{http://dx.doi.org/10.1007/JHEP11(2017)046}{{\em JHEP} {\bfseries 11}
  (2017) 046},
\href{http://arxiv.org/abs/1702.04266}{{\ttfamily arXiv:1702.04266 [hep-th]}}.\, 
A.~Gaikwad, L.~K. Joshi, G.~Mandal, and S.~R. Wadia, ``{Holographic dual to
  charged SYK from 3D Gravity and Chern-Simons},''
  \href{http://arxiv.org/abs/1802.07746}{{\ttfamily arXiv:1802.07746
  [hep-th]}}.



\bibitem{Alekseev:2018pbv}
A.~Alekseev and S.~L. Shatashvili,
  \href{http://dx.doi.org/10.1142/9789813233867_0007}{``{Coadjoint Orbits,
  Cocycles and Gravitational Wess-Zumino},''} in {\em Ludwig Faddeev Memorial
  Volume}, pp.~37--51.
\newblock 2018.
\newblock
\href{http://arxiv.org/abs/1801.07963}{{\ttfamily arXiv:1801.07963 [hep-th]}}.
\newblock




\bibitem{Belokurov:2018aol}
V.~V. Belokurov and E.~T. Shavgulidze, ``{Correlation functions in the
  Schwarzian theory},''
\href{http://arxiv.org/abs/1804.00424}{{\ttfamily arXiv:1804.00424 [hep-th]}}.


\bibitem{GR4}
D.~J. Gross and V.~Rosenhaus, ``{All point correlation functions in SYK},''
  \href{http://dx.doi.org/10.1007/JHEP12(2017)148}{{\em JHEP} {\bfseries 12}
  (2017) 148},
\href{http://arxiv.org/abs/1710.08113}{{\ttfamily arXiv:1710.08113 [hep-th]}}.

\bibitem{GR3}
D.~J. Gross and V.~Rosenhaus, ``{A line of CFTs: from generalized free fields
  to SYK},'' \href{http://dx.doi.org/10.1007/JHEP07(2017)086}{{\em JHEP}
  {\bfseries 07} (2017) 086},
\href{http://arxiv.org/abs/1706.07015}{{\ttfamily arXiv:1706.07015 [hep-th]}}.

\bibitem{Kitaev:2017hnr}
A.~Kitaev, ``{Notes on $\widetilde{\mathrm{SL}}(2,\mathbb{R})$
  representations},''
\href{http://arxiv.org/abs/1711.08169}{{\ttfamily arXiv:1711.08169 [hep-th]}}.

\bibitem{Murugan:2017eto}
J.~Murugan, D.~Stanford, and E.~Witten, ``{More on Supersymmetric and 2d
  Analogs of the SYK Model},''
  \href{http://dx.doi.org/10.1007/JHEP08(2017)146}{{\em JHEP} {\bfseries 08}
  (2017) 146},
\href{http://arxiv.org/abs/1706.05362}{{\ttfamily arXiv:1706.05362 [hep-th]}}.

\bibitem{Bulycheva:2017uqj}
K.~Bulycheva, ``{A note on the SYK model with complex fermions},''
  \href{http://dx.doi.org/10.1007/JHEP12(2017)069}{{\em JHEP} {\bfseries 12}
  (2017) 069},
\href{http://arxiv.org/abs/1706.07411}{{\ttfamily arXiv:1706.07411 [hep-th]}}.

\bibitem{Dobrev:1977qv}
V.~K. Dobrev, G.~Mack, V.~B. Petkova, S.~G. Petrova, and I.~T. Todorov,
  ``{Harmonic Analysis on the n-Dimensional Lorentz Group and Its Application
  to Conformal Quantum Field Theory},''
\href{http://dx.doi.org/10.1007/BFb0009678}{{\em Lect. Notes Phys.} {\bfseries
  63} (1977) 1--280}.

\bibitem{Karateev:2017jgd}
D.~Karateev, P.~Kravchuk, and D.~Simmons-Duffin, ``{Weight Shifting Operators
  and Conformal Blocks},''
  \href{http://dx.doi.org/10.1007/JHEP02(2018)081}{{\em JHEP} {\bfseries 02}
  (2018) 081},
\href{http://arxiv.org/abs/1706.07813}{{\ttfamily arXiv:1706.07813 [hep-th]}}.; 
P.~Kravchuk and D.~Simmons-Duffin, ``{Light-ray operators in conformal field
  theory},''
\href{http://arxiv.org/abs/1805.00098}{{\ttfamily arXiv:1805.00098 [hep-th]}}.

\bibitem{Caron-Huot:2017vep} 
  S.~Caron-Huot,
  ``{Analyticity in Spin in Conformal Theories},''
 \href{https://arxiv.org/abs/1703.00278}{{\ttfamily arXiv:1703.00278 [hep-th]}}.;\, 
  D.~Simmons-Duffin, D.~Stanford and E.~Witten,
  ``{A spacetime derivation of the Lorentzian OPE inversion formula},'' \href{https://arxiv.org/abs/1711.03816}{{\ttfamily arXiv:1711.03816 [hep-th].}}
  

\bibitem{Raben:2018sjl}
T.~Raben and C.-I. Tan, ``{Minkowski Conformal Blocks and the Regge Limit for
  SYK-like Models},''
\href{http://arxiv.org/abs/1801.04208}{{\ttfamily arXiv:1801.04208 [hep-th]}}.

\bibitem{Gadde:2017sjg}
A.~Gadde, ``{In search of conformal theories},''
\href{http://arxiv.org/abs/1702.07362}{{\ttfamily arXiv:1702.07362 [hep-th]}}.

\bibitem{SimmonsDuffin:2012uy}
D.~Simmons-Duffin, ``{Projectors, Shadows, and Conformal Blocks},''
  \href{http://dx.doi.org/10.1007/JHEP04(2014)146}{{\em JHEP} {\bfseries 04}
  (2014) 146},
\href{http://arxiv.org/abs/1204.3894}{{\ttfamily arXiv:1204.3894 [hep-th]}}.

\bibitem{MSS}
J.~Maldacena, S.~H. Shenker, and D.~Stanford, ``{A bound on chaos},''
  \href{http://dx.doi.org/10.1007/JHEP08(2016)106}{{\em JHEP} {\bfseries 08}
  (2016) 106},
\href{http://arxiv.org/abs/1503.01409}{{\ttfamily arXiv:1503.01409 [hep-th]}}.

\bibitem{Grabner:2017pgm} 
  D.~Grabner, N.~Gromov, V.~Kazakov and G.~Korchemsky,
  Phys.\ Rev.\ Lett.\  {\bf 120}, no. 11, 111601 (2018)
  \href{https://arxiv.org/abs/1711.04786}{{\ttfamily arXiv:1711.04786 [hep-th]}}.

\bibitem{Jackiw:1984je}
R.~Jackiw, ``{Lower Dimensional Gravity},''
\href{http://dx.doi.org/10.1016/0550-3213(85)90448-1}{{\em Nucl. Phys.}
  {\bfseries B252} (1985) 343--356}; \, 
C.~Teitelboim, ``{Gravitation and Hamiltonian Structure in Two Space-Time
  Dimensions},''
\href{http://dx.doi.org/10.1016/0370-2693(83)90012-6}{{\em Phys. Lett.}
  {\bfseries B126} (1983) 41--45}.

\bibitem{AP}
A.~Almheiri and J.~Polchinski, ``{Models of AdS$_{2}$ backreaction and
  holography},'' \href{http://dx.doi.org/10.1007/JHEP11(2015)014}{{\em JHEP}
  {\bfseries 11} (2015) 014},
\href{http://arxiv.org/abs/1402.6334}{{\ttfamily arXiv:1402.6334 [hep-th]}}.


\bibitem{Jensen:2016pah}
K.~Jensen, ``{Chaos and hydrodynamics near AdS$_2$},''
  \href{http://dx.doi.org/10.1103/PhysRevLett.117.111601}{{\em Phys. Rev.
  Lett.} {\bfseries 117} no.~11, (2016) 111601},
\href{http://arxiv.org/abs/1605.06098}{{\ttfamily arXiv:1605.06098 [hep-th]}}.

\bibitem{Engelsoy:2016xyb}
J.~Engelsöy, T.~G. Mertens, and H.~Verlinde, ``{An investigation of AdS$_{2}$
  backreaction and holography},''
  \href{http://dx.doi.org/10.1007/JHEP07(2016)139}{{\em JHEP} {\bfseries 07}
  (2016) 139},
\href{http://arxiv.org/abs/1606.03438}{{\ttfamily arXiv:1606.03438 [hep-th]}}.

\bibitem{Dubovsky:2017cnj}
S.~Dubovsky, V.~Gorbenko, and M.~Mirbabayi, ``{Asymptotic fragility, near
  AdS$_{2}$ holography and $ T\overline{T} $},''
  \href{http://dx.doi.org/10.1007/JHEP09(2017)136}{{\em JHEP} {\bfseries 09}
  (2017) 136},
\href{http://arxiv.org/abs/1706.06604}{{\ttfamily arXiv:1706.06604 [hep-th]}}.


\bibitem{Forste:2017apw}
S.~Forste and I.~Golla, ``{Nearly AdS$_2$ sugra and the super-Schwarzian},''
  \href{http://dx.doi.org/10.1016/j.physletb.2017.05.039}{{\em Phys. Lett.}
  {\bfseries B771} (2017) 157--161},
\href{http://arxiv.org/abs/1703.10969}{{\ttfamily arXiv:1703.10969 [hep-th]}};\, 
S.~Forste, J.~Kames-King, and M.~Wiesner, ``{Towards the Holographic Dual of N
  = 2 SYK},'' \href{http://dx.doi.org/10.1007/JHEP03(2018)028}{{\em JHEP}
  {\bfseries 03} (2018) 028},
\href{http://arxiv.org/abs/1712.07398}{{\ttfamily arXiv:1712.07398 [hep-th]}}.





\bibitem{Haehl:2017pak}
F.~M. Haehl and M.~Rozali, ``{Fine Grained Chaos in $AdS_2$ Gravity},''
\href{http://arxiv.org/abs/1712.04963}{{\ttfamily arXiv:1712.04963 [hep-th]}}.


\bibitem{Grumiller:2017qao}
D.~Grumiller, R.~McNees, J.~Salzer, C.~Valcárcel, and D.~Vassilevich,
  ``{Menagerie of AdS$_{2}$ boundary conditions},''
  \href{http://dx.doi.org/10.1007/JHEP10(2017)203}{{\em JHEP} {\bfseries 10}
  (2017) 203},
\href{http://arxiv.org/abs/1708.08471}{{\ttfamily arXiv:1708.08471 [hep-th]}}.



\bibitem{Gonzalez:2018enk}
H.~A. González, D.~Grumiller, and J.~Salzer, ``{Towards a bulk description of
  higher spin SYK},'' \href{http://dx.doi.org/10.1007/JHEP05(2018)083}{{\em
  JHEP} {\bfseries 05} (2018) 083},
\href{http://arxiv.org/abs/1802.01562}{{\ttfamily arXiv:1802.01562 [hep-th]}}.




\bibitem{Lam:2018pvp}
H.~T. Lam, T.~G. Mertens, G.~J. Turiaci, and H.~Verlinde, ``{Shockwave S-matrix
  from Schwarzian Quantum Mechanics},''
\href{http://arxiv.org/abs/1804.09834}{{\ttfamily arXiv:1804.09834 [hep-th]}}.

\bibitem{Qi:2018rqm}
Y.-H. Qi, Y.~Seo, S.-J. Sin, and G.~Song, ``{Schwarzian correction to quantum
  correlation in SYK model},''
\href{http://arxiv.org/abs/1804.06164}{{\ttfamily arXiv:1804.06164 [hep-th]}}.


\bibitem{Taylor:2017dly}
M.~Taylor, ``{Generalized conformal structure, dilaton gravity and SYK},''
  \href{http://dx.doi.org/10.1007/JHEP01(2018)010}{{\em JHEP} {\bfseries 01}
  (2018) 010},
\href{http://arxiv.org/abs/1706.07812}{{\ttfamily arXiv:1706.07812 [hep-th]}}.

\bibitem{Nayak:2018qej}
P.~Nayak, A.~Shukla, R.~M. Soni, S.~P. Trivedi, and V.~Vishal, ``{On the
  Dynamics of Near-Extremal Black Holes},''
\href{http://arxiv.org/abs/1802.09547}{{\ttfamily arXiv:1802.09547 [hep-th]}}.



\bibitem{Li:2018omr}
Y.-Z. Li, S.-L. Li, and H.~Lu, ``{Exact Embeddings of JT Gravity in Strings and
  M-theory},''
\href{http://arxiv.org/abs/1804.09742}{{\ttfamily arXiv:1804.09742 [hep-th]}}.;\, 
I.~Bena, P.~Heidmann, and D.~Turton, ``{AdS$_2$ Holography: Mind the Cap},''
\href{http://arxiv.org/abs/1806.02834}{{\ttfamily arXiv:1806.02834 [hep-th]}}.;\, 
F.~Larsen, ``{A nAttractor Mechanism for nAdS(2)/nCFT(1) Holography},''
\href{http://arxiv.org/abs/1806.06330}{{\ttfamily arXiv:1806.06330 [hep-th]}}.\, 
K.~S. Kolekar and K.~Narayan, ``{$AdS_2$ dilaton gravity from reductions of
  some nonrelativistic theories},''
\href{http://arxiv.org/abs/1803.06827}{{\ttfamily arXiv:1803.06827 [hep-th]}}.


\bibitem{GR2}
D.~J. Gross and V.~Rosenhaus, ``{The Bulk Dual of SYK: Cubic Couplings},''
  \href{http://dx.doi.org/10.1007/JHEP05(2017)092}{{\em JHEP} {\bfseries 05}
  (2017) 092},
\href{http://arxiv.org/abs/1702.08016}{{\ttfamily arXiv:1702.08016 [hep-th]}}.

\bibitem{Das:2017hrt}
S.~R. Das, A.~Ghosh, A.~Jevicki, and K.~Suzuki, ``{Three Dimensional View of
  Arbitrary $q$ SYK models},''
  \href{http://dx.doi.org/10.1007/JHEP02(2018)162}{{\em JHEP} {\bfseries 02}
  (2018) 162},
\href{http://arxiv.org/abs/1711.09839}{{\ttfamily arXiv:1711.09839 [hep-th]}}.

\bibitem{Bardeen:1975gx}
W.~A. Bardeen, I.~Bars, A.~J. Hanson, and R.~D. Peccei, ``{A Study of the
  Longitudinal Kink Modes of the String},''
\href{http://dx.doi.org/10.1103/PhysRevD.13.2364}{{\em Phys. Rev.} {\bfseries
  D13} (1976) 2364--2382}.

\bibitem{GR5}
D.~J. Gross and V.~Rosenhaus{\em {, in progress}} .

\bibitem{Bulycheva:2017ilt}
K.~Bulycheva, I.~R. Klebanov, A.~Milekhin, and G.~Tarnopolsky, ``{Spectra of
  Operators in Large $N$ Tensor Models},''
  \href{http://dx.doi.org/10.1103/PhysRevD.97.026016}{{\em Phys. Rev.}
  {\bfseries D97} no.~2, (2018) 026016},
\href{http://arxiv.org/abs/1707.09347}{{\ttfamily arXiv:1707.09347 [hep-th]}}.

\bibitem{Choudhury:2017tax}
S.~Choudhury, A.~Dey, I.~Halder, L.~Janagal, S.~Minwalla, and R.~Poojary,
  ``{Notes on Melonic $O(N)^{q-1}$ Tensor Models},''
\href{http://arxiv.org/abs/1707.09352}{{\ttfamily arXiv:1707.09352 [hep-th]}}.

\bibitem{Beccaria:2017aqc}
M.~Beccaria and A.~A. Tseytlin, ``{Partition function of free conformal fields
  in 3-plet representation},''
  \href{http://dx.doi.org/10.1007/JHEP05(2017)053}{{\em JHEP} {\bfseries 05}
  (2017) 053},
\href{http://arxiv.org/abs/1703.04460}{{\ttfamily arXiv:1703.04460 [hep-th]}}.

\bibitem{Krishnan:2017lra}
H.~Itoyama, A.~Mironov, and A.~Morozov, ``{Ward identities and combinatorics of
  rainbow tensor models},''
\href{http://arxiv.org/abs/1704.08648}{{\ttfamily arXiv:1704.08648 [hep-th]}}.
P.~Diaz and S.-J. Rey, ``{Orthogonal Bases of Invariants in Tensor Models},''
\href{http://arxiv.org/abs/1706.02667}{{\ttfamily arXiv:1706.02667 [hep-th]}}.;
R.~de~Mello~Koch, R.~Mello~Koch, D.~Gossman, and L.~Tribelhorn, ``{Gauge
  Invariants, Correlators and Holography in Bosonic and Fermionic Tensor
  Models},'' 
\href{http://arxiv.org/abs/1707.01455}{{\ttfamily arXiv:1707.01455 [hep-th]}}.; 
J.~Ben~Geloun and S.~Ramgoolam, ``{Tensor Models, Kronecker coefficients and
  Permutation Centralizer Algebras},''
\href{http://arxiv.org/abs/1708.03524}{{\ttfamily arXiv:1708.03524 [hep-th]}};
C.~Krishnan, K.~V. Pavan~Kumar, and D.~Rosa, ``{Contrasting SYK-like Models},''
\href{http://arxiv.org/abs/1709.06498}{{\ttfamily arXiv:1709.06498 [hep-th]}}.;


\bibitem{Vasiliev:2018zer}
M.~A. Vasiliev, ``{From Coxeter Higher-Spin Theories to Strings and Tensor
  Models},''
\href{http://arxiv.org/abs/1804.06520}{{\ttfamily arXiv:1804.06520 [hep-th]}}.

\bibitem{Maldacena:1997re}
J.~M. Maldacena, ``{The Large N limit of superconformal field theories and
  supergravity},'' \href{http://dx.doi.org/10.1023/A:1026654312961,
  10.4310/ATMP.1998.v2.n2.a1}{{\em Int. J. Theor. Phys.} {\bfseries 38} (1999)
  1113--1133}, \href{http://arxiv.org/abs/hep-th/9711200}{{\ttfamily
  arXiv:hep-th/9711200 [hep-th]}}.
[Adv. Theor. Math. Phys.2,231(1998)].

\bibitem{Gubser:1998bc}
S.~S. Gubser, I.~R. Klebanov, and A.~M. Polyakov, ``{Gauge theory correlators
  from noncritical string theory},''
  \href{http://dx.doi.org/10.1016/S0370-2693(98)00377-3}{{\em Phys. Lett.}
  {\bfseries B428} (1998) 105--114},
\href{http://arxiv.org/abs/hep-th/9802109}{{\ttfamily arXiv:hep-th/9802109
  [hep-th]}}.

\bibitem{Witten:1998qj}
E.~Witten, ``{Anti-de Sitter space and holography},''
  \href{http://dx.doi.org/10.4310/ATMP.1998.v2.n2.a2}{{\em Adv. Theor. Math.
  Phys.} {\bfseries 2} (1998) 253--291},
\href{http://arxiv.org/abs/hep-th/9802150}{{\ttfamily arXiv:hep-th/9802150
  [hep-th]}}.

\bibitem{Klebanov:2002ja}
I.~R. Klebanov and A.~M. Polyakov, ``{AdS dual of the critical O(N) vector
  model},'' \href{http://dx.doi.org/10.1016/S0370-2693(02)02980-5}{{\em Phys.
  Lett.} {\bfseries B550} (2002) 213--219},
\href{http://arxiv.org/abs/hep-th/0210114}{{\ttfamily arXiv:hep-th/0210114
  [hep-th]}}.

\bibitem{Giombi:2016ejx}
S.~Giombi,``{TASI Lectures on the Higher Spin - CFT duality},'' \href{http://arxiv.org/abs/1607.02967}{{\ttfamily arXiv:1607.02967 [hep-th]}}.

\bibitem{Ooguri:2016pdq}
H.~Ooguri and C.~Vafa, ``{Non-supersymmetric AdS and the Swampland},''
\href{http://arxiv.org/abs/1610.01533}{{\ttfamily arXiv:1610.01533 [hep-th]}}.; 
B.~Freivogel and M.~Kleban, ``{Vacua Morghulis},''
\href{http://arxiv.org/abs/1610.04564}{{\ttfamily arXiv:1610.04564 [hep-th]}}.


\bibitem{Cotler:2016fpe}
J.~S. Cotler, G.~Gur-Ari, M.~Hanada, J.~Polchinski, P.~Saad, S.~H. Shenker,
  D.~Stanford, A.~Streicher, and M.~Tezuka, ``{Black Holes and Random
  Matrices},'' \href{http://dx.doi.org/10.1007/JHEP05(2017)118}{{\em JHEP}
  {\bfseries 05} (2017) 118},
\href{http://arxiv.org/abs/1611.04650}{{\ttfamily arXiv:1611.04650 [hep-th]}}\, 

\bibitem{Saad:2018bqo}
P.~Saad, S.~H. Shenker, and D.~Stanford, ``{A semiclassical ramp in SYK and in
  gravity},''
\href{http://arxiv.org/abs/1806.06840}{{\ttfamily arXiv:1806.06840 [hep-th]}}.

\bibitem{Maldacena:2017axo}
J.~Maldacena, D.~Stanford, and Z.~Yang, ``{Diving into traversable
  wormholes},'' \href{http://dx.doi.org/10.1002/prop.201700034}{{\em Fortsch.
  Phys.} {\bfseries 65} no.~5, (2017) 1700034},
\href{http://arxiv.org/abs/1704.05333}{{\ttfamily arXiv:1704.05333 [hep-th]}}.

\bibitem{Kourkoulou:2017zaj}
I.~Kourkoulou and J.~Maldacena, ``{Pure states in the SYK model and
  nearly-$AdS_2$ gravity},''
\href{http://arxiv.org/abs/1707.02325}{{\ttfamily arXiv:1707.02325 [hep-th]}}.


\bibitem{Maldacena:2018lmt}
J.~Maldacena and X.-L. Qi, ``{Eternal traversable wormhole},''
\href{http://arxiv.org/abs/1804.00491}{{\ttfamily arXiv:1804.00491 [hep-th]}}.



\bibitem{Brown:2017jil}
A.~R. Brown and L.~Susskind, ``{Second law of quantum complexity},''
\href{http://arxiv.org/abs/1701.01107}{{\ttfamily arXiv:1701.01107 [hep-th]}}.


\bibitem{Caputa:2017yrh}
P.~Caputa, N.~Kundu, M.~Miyaji, T.~Takayanagi, and K.~Watanabe, ``{Liouville
  Action as Path-Integral Complexity: From Continuous Tensor Networks to
  AdS/CFT},'' \href{http://dx.doi.org/10.1007/JHEP11(2017)097}{{\em JHEP}
  {\bfseries 11} (2017) 097},
\href{http://arxiv.org/abs/1706.07056}{{\ttfamily arXiv:1706.07056 [hep-th]}}.

\bibitem{Harlow:2018tqv}
D.~Harlow and D.~Jafferis, ``{The Factorization Problem in Jackiw-Teitelboim
  Gravity},''
\href{http://arxiv.org/abs/1804.01081}{{\ttfamily arXiv:1804.01081 [hep-th]}}.

\bibitem{Swingle}
B.~Swingle, ``Spacetime from entanglement,''
  \href{http://dx.doi.org/10.1146/annurev-conmatphys-033117-054219}{{\em Annual
  Review of Condensed Matter Physics} {\bfseries 9} no.~1, (2018) 345--358},
  \url{https://doi.org/10.1146/annurev-conmatphys-033117-054219}.





\bibitem{Fu:2016vas}
W.~Fu, D.~Gaiotto, J.~Maldacena, and S.~Sachdev, ``{Supersymmetric
  Sachdev-Ye-Kitaev models},''
  \href{http://dx.doi.org/10.1103/PhysRevD.95.069904,
  10.1103/PhysRevD.95.026009}{{\em Phys. Rev.} {\bfseries D95} no.~2, (2017)
  026009}, \href{http://arxiv.org/abs/1610.08917}{{\ttfamily arXiv:1610.08917
  [hep-th]}}.


\bibitem{Sannomiya:2016mnj}
N.~Sannomiya, H.~Katsura, and Y.~Nakayama, ``{Supersymmetry breaking and
  Nambu-Goldstone fermions with cubic dispersion},''
  \href{http://dx.doi.org/10.1103/PhysRevD.95.065001}{{\em Phys. Rev.}
  {\bfseries D95} no.~6, (2017) 065001},
\href{http://arxiv.org/abs/1612.02285}{{\ttfamily arXiv:1612.02285
  [cond-mat.str-el]}}.


\bibitem{Peng:2017spg}
C.~Peng, M.~Spradlin, and A.~Volovich, ``{Correlators in the $\mathcal{N}=2$
  Supersymmetric SYK Model},''
  \href{http://dx.doi.org/10.1007/JHEP10(2017)202}{{\em JHEP} {\bfseries 10}
  (2017) 202},
\href{http://arxiv.org/abs/1706.06078}{{\ttfamily arXiv:1706.06078 [hep-th]}}.


\bibitem{Bulycheva:2018qcp}
K.~Bulycheva, ``{$ \mathcal{N}=2 $ SYK model in the superspace formalism},''
  \href{http://dx.doi.org/10.1007/JHEP04(2018)036}{{\em JHEP} {\bfseries 04}
  (2018) 036},
\href{http://arxiv.org/abs/1801.09006}{{\ttfamily arXiv:1801.09006 [hep-th]}}.

\bibitem{Narayan:2017hvh}
P.~Narayan and J.~Yoon, ``{Supersymmetric SYK Model with Global Symmetry},''
\href{http://arxiv.org/abs/1712.02647}{{\ttfamily arXiv:1712.02647 [hep-th]}}.


\bibitem{Anninos:2016szt}
D.~Anninos, T.~Anous, and F.~Denef, ``{Disordered Quivers and Cold Horizons},''
  \href{http://dx.doi.org/10.1007/JHEP12(2016)071}{{\em JHEP} {\bfseries 12}
  (2016) 071},
\href{http://arxiv.org/abs/1603.00453}{{\ttfamily arXiv:1603.00453 [hep-th]}}.

\bibitem{Giombi:2017dtl}
S.~Giombi, I.~R. Klebanov, and G.~Tarnopolsky, ``{Bosonic tensor models at
  large $N$ and small $\epsilon$},''
  \href{http://dx.doi.org/10.1103/PhysRevD.96.106014}{{\em Phys. Rev.}
  {\bfseries D96} no.~10, (2017) 106014},
\href{http://arxiv.org/abs/1707.03866}{{\ttfamily arXiv:1707.03866 [hep-th]}}.


\bibitem{Prakash:2017hwq}
S.~Prakash and R.~Sinha, ``{A Complex Fermionic Tensor Model in $d$
  Dimensions},'' \href{http://dx.doi.org/10.1007/JHEP02(2018)086}{{\em JHEP}
  {\bfseries 02} (2018) 086},
\href{http://arxiv.org/abs/1710.09357}{{\ttfamily arXiv:1710.09357 [hep-th]}}.


\bibitem{Benedetti:2017fmp}
D.~Benedetti, S.~Carrozza, R.~Gurau, and A.~Sfondrini, ``{Tensorial Gross-Neveu
  models},'' \href{http://dx.doi.org/10.1007/JHEP01(2018)003}{{\em JHEP}
  {\bfseries 01} (2018) 003},
\href{http://arxiv.org/abs/1710.10253}{{\ttfamily arXiv:1710.10253 [hep-th]}}.



\bibitem{Azeyanagi:2017drg}
T.~Azeyanagi, F.~Ferrari, and F.~I. Schaposnik~Massolo, ``{Phase Diagram of
  Planar Matrix Quantum Mechanics, Tensor, and Sachdev-Ye-Kitaev Models},''
  \href{http://dx.doi.org/10.1103/PhysRevLett.120.061602}{{\em Phys. Rev.
  Lett.} {\bfseries 120} no.~6, (2018) 061602},
\href{http://arxiv.org/abs/1707.03431}{{\ttfamily arXiv:1707.03431 [hep-th]}}.


\bibitem{Peng:2016mxj}
C.~Peng, M.~Spradlin, and A.~Volovich, ``{A Supersymmetric SYK-like Tensor
  Model},'' \href{http://dx.doi.org/10.1007/JHEP05(2017)062}{{\em JHEP}
  {\bfseries 05} (2017) 062},
\href{http://arxiv.org/abs/1612.03851}{{\ttfamily arXiv:1612.03851 [hep-th]}}.



\bibitem{Chang:2018sve}
C.-M. Chang, S.~Colin-Ellerin, and M.~Rangamani, ``{On Melonic Supertensor
  Models},''
\href{http://arxiv.org/abs/1806.09903}{{\ttfamily arXiv:1806.09903 [hep-th]}}.



\bibitem{Bi:2017yvx}
Z.~Bi, C.-M. Jian, Y.-Z. You, K.~A. Pawlak, and C.~Xu, ``{Instability of the
  non-Fermi liquid state of the Sachdev-Ye-Kitaev Model},''
  \href{http://dx.doi.org/10.1103/PhysRevB.95.205105}{{\em Phys. Rev.}
  {\bfseries B95} no.~20, (2017) 205105},
\href{http://arxiv.org/abs/1701.07081}{{\ttfamily arXiv:1701.07081
  [cond-mat.str-el]}}.


\bibitem{Berkooz:2016cvq}
M.~Berkooz, P.~Narayan, M.~Rozali, and J.~Simon, ``{Higher Dimensional
  Generalizations of the SYK Model},''
  \href{http://dx.doi.org/10.1007/JHEP01(2017)138}{{\em JHEP} {\bfseries 01}
  (2017) 138},
\href{http://arxiv.org/abs/1610.02422}{{\ttfamily arXiv:1610.02422 [hep-th]}};\, 
M.~Berkooz, P.~Narayan, M.~Rozali, and J.~Simon, ``{Comments on the Random
  Thirring Model},'' \href{http://dx.doi.org/10.1007/JHEP09(2017)057}{{\em
  JHEP} {\bfseries 09} (2017) 057},
\href{http://arxiv.org/abs/1702.05105}{{\ttfamily arXiv:1702.05105 [hep-th]}}.



\bibitem{Peng:2018zap}
C.~Peng, ``{$\mathcal{N}=(0,2)$ SYK, Chaos and Higher-Spins},''
\href{http://arxiv.org/abs/1805.09325}{{\ttfamily arXiv:1805.09325 [hep-th]}}.


\bibitem{Turiaci:2017zwd}
G.~Turiaci and H.~Verlinde, ``{Towards a 2d QFT Analog of the SYK Model},''
  \href{http://dx.doi.org/10.1007/JHEP10(2017)167}{{\em JHEP} {\bfseries 10}
  (2017) 167},
\href{http://arxiv.org/abs/1701.00528}{{\ttfamily arXiv:1701.00528 [hep-th]}}.









\bibitem{Gubser:2017qed}
S.~S. Gubser, M.~Heydeman, C.~Jepsen, S.~Parikh, I.~Saberi, B.~Stoica, and
  B.~Trundy, ``{Signs of the time: Melonic theories over diverse number
  systems},''
\href{http://arxiv.org/abs/1707.01087}{{\ttfamily arXiv:1707.01087 [hep-th]}}.



\bibitem{Gu:2016oyy}
Y.~Gu, X.-L. Qi, and D.~Stanford, ``{Local criticality, diffusion and chaos in
  generalized Sachdev-Ye-Kitaev models},''
  \href{http://dx.doi.org/10.1007/JHEP05(2017)125}{{\em JHEP} {\bfseries 05}
  (2017) 125}, \href{http://arxiv.org/abs/1609.07832}{{\ttfamily
  arXiv:1609.07832}}.

\bibitem{Song:2017pfw}
X.-Y. {Song}, C.-M. {Jian}, and L.~{Balents}, ``{Strongly Correlated Metal
  Built from Sachdev-Ye-Kitaev Models},''
  \href{http://dx.doi.org/10.1103/PhysRevLett.119.216601}{{\em Phys. Rev.
  Lett.} {\bfseries 119} no.~21, (Nov., 2017) 216601},
  \href{http://arxiv.org/abs/1705.00117}{{\ttfamily arXiv:1705.00117
  [cond-mat.str-el]}}.
  
  \bibitem{Banerjee:2016ncu}
S.~Banerjee and E.~Altman, ``{Solvable model for a dynamical quantum phase
  transition from fast to slow scrambling},''
  \href{http://dx.doi.org/10.1103/PhysRevB.95.134302}{{\em Phys. Rev.}
  {\bfseries B95} no.~13, (2017) 134302},
\href{http://arxiv.org/abs/1610.04619}{{\ttfamily arXiv:1610.04619
  [cond-mat.str-el]}}.




\bibitem{Jian:2017unn}
S.-K. Jian and H.~Yao, ``{Solvable Sachdev-Ye-Kitaev models in higher
  dimensions: from diffusion to many-body localization},''
  \href{http://dx.doi.org/10.1103/PhysRevLett.119.206602}{{\em Phys. Rev.
  Lett.} {\bfseries 119} no.~20, (2017) 206602},
\href{http://arxiv.org/abs/1703.02051}{{\ttfamily arXiv:1703.02051
  [cond-mat.str-el]}}.





\bibitem{Chowdhury:2018sho}
D.~Chowdhury, Y.~Werman, E.~Berg, and T.~Senthil, ``{Translationally invariant
  non-Fermi liquid metals with critical Fermi-surfaces: Solvable models},''
\href{http://arxiv.org/abs/1801.06178}{{\ttfamily arXiv:1801.06178
  [cond-mat.str-el]}}.



\bibitem{Fu:2018spl}
W.~Fu, Y.~Gu, S.~Sachdev, and G.~Tarnopolsky, ``{$\mathbb Z_2$ fractionalized
  phases of a solvable, disordered, $t$-$J$ model},''
\href{http://arxiv.org/abs/1804.04130}{{\ttfamily arXiv:1804.04130
  [cond-mat.str-el]}}.

\bibitem{Patel:2018rjp} 
  A.~A.~Patel, M.~J.~Lawler and E.~A.~Kim,
  ``{Coherent superconductivity with large gap ratio from incoherent metals,}''
  \href{https://arxiv.org/abs/1805.11098}{{\ttfamily  arXiv:1805.11098 [cond-mat.str-el]}}.

\bibitem{Wu:2018vua}
X.~Wu, X.~Chen, C.-M. Jian, Y.-Z. You, and C.~Xu, ``{A candidate Theory for the
  "Strange Metal" phase at Finite Energy Window},''
\href{http://arxiv.org/abs/1802.04293}{{\ttfamily arXiv:1802.04293
  [cond-mat.str-el]}}.


\bibitem{Davison:2016ngz}
R.~A. Davison, W.~Fu, A.~Georges, Y.~Gu, K.~Jensen, and S.~Sachdev,
  ``{Thermoelectric transport in disordered metals without quasiparticles: The
  Sachdev-Ye-Kitaev models and holography},''
  \href{http://dx.doi.org/10.1103/PhysRevB.95.155131}{{\em Phys. Rev.}
  {\bfseries B95} no.~15, (2017) 155131},
\href{http://arxiv.org/abs/1612.00849}{{\ttfamily arXiv:1612.00849
  [cond-mat.str-el]}}.


\bibitem{Gu:2017ohj}
Y.~Gu, A.~Lucas, and X.-L. Qi, ``{Energy diffusion and the butterfly effect in
  inhomogeneous Sachdev-Ye-Kitaev chains},''
  \href{http://dx.doi.org/10.21468/SciPostPhys.2.3.018}{{\em SciPost Phys.}
  {\bfseries 2} no.~3, (2017) 018},
\href{http://arxiv.org/abs/1702.08462}{{\ttfamily arXiv:1702.08462 [hep-th]}}.

\bibitem{Peng:2017kro}
C.~Peng, ``{Vector models and generalized SYK models},''
  \href{http://dx.doi.org/10.1007/JHEP05(2017)129}{{\em JHEP} {\bfseries 05}
  (2017) 129},
\href{http://arxiv.org/abs/1704.04223}{{\ttfamily arXiv:1704.04223 [hep-th]}}.




\bibitem{Chen:2017dbb}
Y.~Chen, H.~Zhai, and P.~Zhang, ``{Tunable Quantum Chaos in the
  Sachdev-Ye-Kitaev Model Coupled to a Thermal Bath},''
  \href{http://dx.doi.org/10.1007/JHEP07(2017)150}{{\em JHEP} {\bfseries 07}
  (2017) 150},
\href{http://arxiv.org/abs/1705.09818}{{\ttfamily arXiv:1705.09818 [hep-th]}}.

\bibitem{Patel:2017mjv}
A.~A. Patel, J.~McGreevy, D.~P. Arovas, and S.~Sachdev, ``{Magnetotransport in
  a model of a disordered strange metal},''
  \href{http://dx.doi.org/10.1103/PhysRevX.8.021049}{{\em Phys. Rev.}
  {\bfseries X8} no.~2, (2018) 021049},
\href{http://arxiv.org/abs/1712.05026}{{\ttfamily arXiv:1712.05026
  [cond-mat.str-el]}}.

\bibitem{Mondal:2018xwy}
S.~Mondal, ``{Chaos, Instability and a Stranger Metal},''
\href{http://arxiv.org/abs/1801.09669}{{\ttfamily arXiv:1801.09669 [hep-th]}}.


\bibitem{Blake:2016jnn}
M.~Blake and A.~Donos, ``{Diffusion and Chaos from near AdS$_2$ horizons},''
  \href{http://dx.doi.org/10.1007/JHEP02(2017)013}{{\em JHEP} {\bfseries 02}
  (2017) 013},
\href{http://arxiv.org/abs/1611.09380}{{\ttfamily arXiv:1611.09380 [hep-th]}}.

\bibitem{Kukuljan:2017xag}
I.~Kukuljan, S.~Grozdanov, and T.~Prosen, ``{Weak Quantum Chaos},''
  \href{http://dx.doi.org/10.1103/PhysRevB.96.060301}{{\em Phys. Rev.}
  {\bfseries B96} no.~6, (2017) 060301},
\href{http://arxiv.org/abs/1701.09147}{{\ttfamily arXiv:1701.09147
  [cond-mat.stat-mech]}}.



\bibitem{Scaffidi:2017ghs}
T.~Scaffidi and E.~Altman, ``{Semiclassical Theory of Many-Body Quantum Chaos
  and its Bound},''
\href{http://arxiv.org/abs/1711.04768}{{\ttfamily arXiv:1711.04768
  [cond-mat.stat-mech]}}.



\bibitem{Liu:2017kfa}
C.~Liu, X.~Chen, and L.~Balents, ``{Quantum Entanglement of the
  Sachdev-Ye-Kitaev Models},''
  \href{http://dx.doi.org/10.1103/PhysRevB.97.245126}{{\em Phys. Rev.}
  {\bfseries B97} no.~24, (2018) 245126},
\href{http://arxiv.org/abs/1709.06259}{{\ttfamily arXiv:1709.06259
  [cond-mat.str-el]}}.

\bibitem{Gu:2017njx}
Y.~Gu, A.~Lucas, and X.-L. Qi, ``{Spread of entanglement in a Sachdev-Ye-Kitaev
  chain},'' \href{http://dx.doi.org/10.1007/JHEP09(2017)120}{{\em JHEP}
  {\bfseries 09} (2017) 120},
\href{http://arxiv.org/abs/1708.00871}{{\ttfamily arXiv:1708.00871 [hep-th]}}.


\bibitem{Eberlein:2017wah}
A.~Eberlein, V.~Kasper, S.~Sachdev, and J.~Steinberg, ``{Quantum quench of the
  Sachdev-Ye-Kitaev Model},''
  \href{http://dx.doi.org/10.1103/PhysRevB.96.205123}{{\em Phys. Rev.}
  {\bfseries B96} no.~20, (2017) 205123},
\href{http://arxiv.org/abs/1706.07803}{{\ttfamily arXiv:1706.07803
  [cond-mat.str-el]}}.

\bibitem{Huang:2017nox}
Y.~Huang and Y.~Gu, ``{Eigenstate entanglement in the Sachdev-Ye-Kitaev
  model},''
\href{http://arxiv.org/abs/1709.09160}{{\ttfamily arXiv:1709.09160 [hep-th]}}.

\bibitem{Sonner:2017hxc}
J.~Sonner and M.~Vielma, ``{Eigenstate thermalization in the Sachdev-Ye-Kitaev
  model},'' \href{http://dx.doi.org/10.1007/JHEP11(2017)149}{{\em JHEP}
  {\bfseries 11} (2017) 149},
\href{http://arxiv.org/abs/1707.08013}{{\ttfamily arXiv:1707.08013 [hep-th]}}.

\bibitem{Haque:2017bts}
M.~Haque and P.~McClarty, ``{Eigenstate Thermalization Scaling in Majorana
  Clusters: from Integrable to Chaotic SYK Models},''
\href{http://arxiv.org/abs/1711.02360}{{\ttfamily arXiv:1711.02360
  [cond-mat.stat-mech]}}.

\bibitem{Hunter-Jones:2017raw}
N.~Hunter-Jones, J.~Liu, and Y.~Zhou, ``{On thermalization in the SYK and
  supersymmetric SYK models},''
  \href{http://dx.doi.org/10.1007/JHEP02(2018)142}{{\em JHEP} {\bfseries 02}
  (2018) 142},
\href{http://arxiv.org/abs/1710.03012}{{\ttfamily arXiv:1710.03012 [hep-th]}}.

\bibitem{Danshita:2016xbo}
I.~Danshita, M.~Hanada, and M.~Tezuka, ``{Creating and probing the
  Sachdev-Ye-Kitaev model with ultracold gases: Towards experimental studies of
  quantum gravity},'' \href{http://dx.doi.org/10.1093/ptep/ptx108}{{\em PTEP}
  {\bfseries 2017} no.~8, (2017) 083I01},
\href{http://arxiv.org/abs/1606.02454}{{\ttfamily arXiv:1606.02454
  [cond-mat.quant-gas]}}.

\bibitem{Garcia-Alvarez:2016wem}
L.~García-Álvarez, I.~L. Egusquiza, L.~Lamata, A.~del Campo, J.~Sonner, and
  E.~Solano, ``{Digital Quantum Simulation of Minimal AdS/CFT},''
  \href{http://dx.doi.org/10.1103/PhysRevLett.119.040501}{{\em Phys. Rev.
  Lett.} {\bfseries 119} no.~4, (2017) 040501},
\href{http://arxiv.org/abs/1607.08560}{{\ttfamily arXiv:1607.08560
  [quant-ph]}}.





\bibitem{Pikulin:2017mhj}
D.~I. Pikulin and M.~Franz, ``{Black Hole on a Chip: Proposal for a Physical
  Realization of the Sachdev-Ye-Kitaev model in a Solid-State System},''
  \href{http://dx.doi.org/10.1103/PhysRevX.7.031006}{{\em Phys. Rev.}
  {\bfseries X7} no.~3, (2017) 031006},
\href{http://arxiv.org/abs/1702.04426}{{\ttfamily arXiv:1702.04426
  [cond-mat.dis-nn]}}.



\bibitem{Chew:2017xuo}
A.~Chew, A.~Essin, and J.~Alicea, ``{Approximating the Sachdev-Ye-Kitaev model
  with Majorana wires},''
  \href{http://dx.doi.org/10.1103/PhysRevB.96.121119}{{\em Phys. Rev.}
  {\bfseries B96} no.~12, (2017) 121119},
\href{http://arxiv.org/abs/1703.06890}{{\ttfamily arXiv:1703.06890
  [cond-mat.dis-nn]}}.


\bibitem{Luo:2017bno}
Z.~Luo, Y.-Z. You, J.~Li, C.-M. Jian, D.~Lu, C.~Xu, B.~Zeng, and R.~Laflamme,
  ``{Observing Fermion Pair Instability of the Sachdev-Ye-Kitaev Model on a
  Quantum Spin Simulator},''
\href{http://arxiv.org/abs/1712.06458}{{\ttfamily arXiv:1712.06458
  [quant-ph]}}.

\bibitem{Chen:2018wvb}
A.~Chen, R.~Ilan, F.~de~Juan, D.~I. Pikulin, and M.~Franz, ``{Quantum
  holography in a graphene flake with an irregular boundary},''
\href{http://arxiv.org/abs/1802.00802}{{\ttfamily arXiv:1802.00802
  [cond-mat.str-el]}}.

\bibitem{2018arXiv180602793B}
R. Babbush, D.Berry, and H. Neven, 
 ``{Quantum Simulation of the Sachdev-Ye-Kitaev Model by Asymmetric Qubitization}",
\href{https://arxiv.org/abs/1806.02793}{{\ttfamily arXiv:1806.02793 [quant-ph]}}.



\bibitem{Parcollet:1997ysb}
O.~Parcollet, A.~Georges, G.~Kotliar, and A.~Sengupta, ``{Overscreened
  multichannel SU(N) Kondo model: Large-N solution and conformal field
  theory},'' \href{http://dx.doi.org/10.1103/PhysRevB.58.3794}{{\em Phys. Rev.}
  {\bfseries B58} no.~7, (1998) 3794},
\href{http://arxiv.org/abs/cond-mat/9711192}
{{\ttfamily arXiv:cond-mat/9711192
}}.

\bibitem{Sachdev:2010um}
S.~Sachdev, ``{Holographic metals and the fractionalized Fermi liquid},''
  \href{http://dx.doi.org/10.1103/PhysRevLett.105.151602}{{\em Phys. Rev.
  Lett.} {\bfseries 105} (2010) 151602},
\href{http://arxiv.org/abs/1006.3794}{{\ttfamily arXiv:1006.3794 [hep-th]}}.

\bibitem{Sachdev:2015efa}
S.~Sachdev, ``{Bekenstein-Hawking Entropy and Strange Metals},''
  \href{http://dx.doi.org/10.1103/PhysRevX.5.041025}{{\em Phys. Rev.}
  {\bfseries X5} no.~4, (2015) 041025},
\href{http://arxiv.org/abs/1506.05111}{{\ttfamily arXiv:1506.05111 [hep-th]}}.






\bibitem{Garcia-Garcia:2016mno}
A.~M. García-García and J.~J.~M. Verbaarschot, ``{Spectral and thermodynamic
  properties of the Sachdev-Ye-Kitaev model},''
  \href{http://dx.doi.org/10.1103/PhysRevD.94.126010}{{\em Phys. Rev.}
  {\bfseries D94} no.~12, (2016) 126010},
\href{http://arxiv.org/abs/1610.03816}{{\ttfamily arXiv:1610.03816 [hep-th]}}; 
A.~M. García-García and J.~J.~M. Verbaarschot, ``{Analytical Spectral Density
  of the Sachdev-Ye-Kitaev Model at finite N},''
  \href{http://dx.doi.org/10.1103/PhysRevD.96.066012}{{\em Phys. Rev.}
  {\bfseries D96} no.~6, (2017) 066012},
\href{http://arxiv.org/abs/1701.06593}{{\ttfamily arXiv:1701.06593 [hep-th]}};\, 
Y.~Jia and J.~J.~M. Verbaarschot, ``{Large $N$ expansion of the moments and
  free energy of Sachdev-Ye-Kitaev model, and the enumeration of intersection
  graphs},''
\href{http://arxiv.org/abs/1806.03271}{{\ttfamily arXiv:1806.03271 [hep-th]}};\, 

\bibitem{You:2016ldz}
Y.-Z. You, A.~W.~W. Ludwig, and C.~Xu, ``{Sachdev-Ye-Kitaev Model and
  Thermalization on the Boundary of Many-Body Localized Fermionic Symmetry
  Protected Topological States},''
  \href{http://dx.doi.org/10.1103/PhysRevB.95.115150}{{\em Phys. Rev.}
  {\bfseries B95} no.~11, (2017) 115150},
\href{http://arxiv.org/abs/1602.06964}{{\ttfamily arXiv:1602.06964
  [cond-mat.str-el]}}.





\bibitem{Krishnan:2016bvg}
C.~Krishnan, S.~Sanyal, and P.~N. Bala~Subramanian, ``{Quantum Chaos and
  Holographic Tensor Models},''
  \href{http://dx.doi.org/10.1007/JHEP03(2017)056}{{\em JHEP} {\bfseries 03}
  (2017) 056},
\href{http://arxiv.org/abs/1612.06330}{{\ttfamily arXiv:1612.06330 [hep-th]}}; 
C.~Krishnan, K.~V.~P. Kumar, and S.~Sanyal, ``{Random Matrices and Holographic
  Tensor Models},'' \href{http://dx.doi.org/10.1007/JHEP06(2017)036}{{\em JHEP}
  {\bfseries 06} (2017) 036},
\href{http://arxiv.org/abs/1703.08155}{{\ttfamily arXiv:1703.08155 [hep-th]}}.

\bibitem{Liu:2016rdi}
Y.~Liu, M.~A. Nowak, and I.~Zahed, ``{Disorder in the Sachdev-Yee-Kitaev
  Model},'' \href{http://dx.doi.org/10.1016/j.physletb.2017.08.054}{{\em Phys.
  Lett.} {\bfseries B773} (2017) 647--653},
\href{http://arxiv.org/abs/1612.05233}{{\ttfamily arXiv:1612.05233 [hep-th]}}.


\bibitem{Li:2017hdt}
T.~Li, J.~Liu, Y.~Xin, and Y.~Zhou, ``{Supersymmetric SYK model and random
  matrix theory},'' \href{http://dx.doi.org/10.1007/JHEP06(2017)111}{{\em JHEP}
  {\bfseries 06} (2017) 111},
\href{http://arxiv.org/abs/1702.01738}{{\ttfamily arXiv:1702.01738 [hep-th]}}.




\bibitem{Chaudhuri:2017vrv}
S.~Chaudhuri, V.~I. Giraldo-Rivera, A.~Joseph, R.~Loganayagam, and J.~Yoon,
  ``{Abelian Tensor Models on the Lattice},''
  \href{http://dx.doi.org/10.1103/PhysRevD.97.086007}{{\em Phys. Rev.}
  {\bfseries D97} no.~8, (2018) 086007},
\href{http://arxiv.org/abs/1705.01930}{{\ttfamily arXiv:1705.01930 [hep-th]}}.


\bibitem{Kanazawa:2017dpd}
T.~Kanazawa and T.~Wettig, ``{Complete random matrix classification of SYK
  models with $\mathcal{N}=0$, $1$ and $2$ supersymmetry},''
  \href{http://dx.doi.org/10.1007/JHEP09(2017)050}{{\em JHEP} {\bfseries 09}
  (2017) 050},
\href{http://arxiv.org/abs/1706.03044}{{\ttfamily arXiv:1706.03044 [hep-th]}}.

\bibitem{Garcia-Garcia:2017bkg} 
  A.~M.~García-García, B.~Loureiro, A.~Romero-Bermúdez and M.~Tezuka,
  `{`Chaotic-Integrable Transition in the Sachdev-Ye-Kitaev Model},''
 \href{https://arxiv.org/abs/1707.02197}{{\ttfamily   arXiv:1707.02197 [hep-th]}}.
  
  \bibitem{Kos:2017zjh} 
  P.~Kos, M.~Ljubotina and T.~Prosen,
  ``{Many-body quantum chaos: Analytic connection to random matrix theory},''
 {\em Phys.\ Rev.\ X} {\bf 8}, no. 2, 021062 (2018)
 \href{https://arxiv.org/abs/1712.02665}{{\ttfamily arXiv:1712.02665 [nlin.CD]}}.

\bibitem{Altland:2017eao}
A.~Altland and D.~Bagrets, ``{Quantum ergodicity in the SYK model},''
  \href{http://dx.doi.org/10.1016/j.nuclphysb.2018.02.015}{{\em Nucl. Phys.}
  {\bfseries B930} (2018) 45--68},
\href{http://arxiv.org/abs/1712.05073}{{\ttfamily arXiv:1712.05073
  [cond-mat.str-el]}}.


\bibitem{Feng:2018zsx}
R.~Feng, G.~Tian, and D.~Wei, ``{Spectrum of SYK model},''
\href{http://arxiv.org/abs/1801.10073}{{\ttfamily arXiv:1801.10073 [math-ph]}}.















\end{thebibliography}

}
\end{changemargin}

\end{document}